\documentclass[12pt]{article}

\usepackage{amssymb}

\usepackage{svg}
\usepackage{amsmath}
\usepackage{float}
\usepackage{pdflscape}
\usepackage{setspace}
\usepackage{float}
\usepackage[caption = false]{subfig}
\usepackage[export]{adjustbox}
\usepackage{wrapfig}
\usepackage[section]{placeins}
\usepackage[margin=1cm]{geometry}
\usepackage{caption}
\usepackage{tabularx}
\usepackage{amsmath,mathabx}
\usepackage{mathtools, nccmath} 
\usepackage{afterpage} 
\usepackage{tikz} 

\newlength\replength
\newcommand\repfrac{.40}
\newcommand\rulewidth{.6pt}

\makeatletter
\newcommand*\dashline{$\dabar@\dabar@\dabar@$}

\def\dashht{.5\dimexpr\ht\strutbox-\dp\strutbox\relax}
\newcommand\tdashfill[1][\repfrac]{\cleaders\hbox to \replength{%
  \smash{\rule[\dashht]{\repfrac\replength}{\rulewidth}%
  \kern.5\dimexpr\replength-\repfrac\replength-2.5pt\relax%
  \raisebox{\dimexpr\dashht-.3pt}{.}}}\hfill}
\newcommand\dashdotline[1]{%
  \makebox[#1][l]{\tdashfill\hfil}}

\renewcommand{\thefootnote}{\alph{footnote}}

\newcommand{\astfootnote}[1]{%
\let\oldthefootnote=\thefootnote%
\setcounter{footnote}{0}%
\renewcommand{\thefootnote}{\fnsymbol{footnote}}%
\footnote{#1}%
\let\thefootnote=\oldthefootnote%
}

\newcommand{\beginsupplement}{%
        \setcounter{table}{0}
        \renewcommand{\thetable}{S\arabic{table}}%
        \setcounter{figure}{0}
        \renewcommand{\thefigure}{S\arabic{figure}}%
     }

\usepackage{graphicx,scalerel}
\newcommand\sbullet[1][.5]{\mathbin{\ThisStyle{\vcenter{\hbox{%
  \scalebox{#1}{$\SavedStyle\bullet$}}}}}%
}

\usepackage[english]{babel}
\usepackage[hidelinks]{hyperref}
\usepackage{xcolor}

\usepackage{microtype}

\definecolor{darkorange}{RGB}{221,110,78}
\definecolor{darkred}{RGB}{153,51,51}
\definecolor{darkgray}{RGB}{77,75,75}
\definecolor{midblue}{RGB}{1,128,255}
\definecolor{darkblue}{RGB}{0,51,102}
\definecolor{lime}{RGB}{0,233 ,0}
\definecolor{greenblue}{RGB}{0,142,100}
\definecolor{darkgreen}{RGB}{0,102,51}
\definecolor{midgreen}{RGB}{76,153,0}
\definecolor{dead}{RGB}{30,144,255}
\definecolor{fix}{RGB}{50,205,50}
\definecolor{cyclic}{RGB}{255,215,0}
\definecolor{chaos}{RGB}{255,69,0}
\definecolor{ns}{RGB}{139,69,19}
\definecolor{classicgreen}{RGB}{0, 128, 0}
\definecolor{classicred}{RGB}{255, 0, 0}
\definecolor{classicorange}{RGB}{255, 165, 0}
\definecolor{firebrick}{RGB}{178, 34, 34}
\definecolor{black}{RGB}{0, 0, 0}

\usepackage[font=footnotesize,labelfont={bf,sc,small},figurename=Fig.,tablename=Tab.]{caption}
\usepackage[labelsep=endash]{caption}

\usepackage{times}

\newenvironment{sciabstract}{%
\begin{quote} \bf}
{\end{quote}}

\newcounter{lastnote}

\title{Excitatory/Inhibitory Balance Emerges as a Key Factor for RBN Performance, Overriding Attractor Dynamics}

\author
{Emmanuel Calvet,$^{1\ast}$ Jean Rouat,$^{1}$ Bertrand Reulet$^{2}$\\
\\
\normalsize{$^{1}$\href{https://www.gegi.usherbrooke.ca/necotis/?lang=en}{NECOTIS}, Génie électrique, Université de Sherbrooke, Canada}\\
\normalsize{$^{2}$\href{https://www.usherbrooke.ca/iq/en/}{Institut Quantique}, Département de Physique, Université de Sherbrooke, Canada}\\
\\
\normalsize{$^\ast$E-mail:  emmanuel.calvet@usherbrooke.ca}
}

\date{}

\begin{document} 
\setlength{\belowdisplayskip}{0.1pt} \setlength{\belowdisplayshortskip}{0.1pt}
\setlength{\abovedisplayskip}{0pt} \setlength{\abovedisplayshortskip}{0pt}
\appto{\bibsetup}{\raggedright}


\baselineskip24pt


\maketitle

\setstretch{1.20}

\begin{sciabstract}
Reservoir computing provides a time and cost-efficient alternative to traditional learning methods. Critical regimes, known as the "edge of chaos," have been found to optimize computational performance in binary neural networks. However, little attention has been devoted to studying reservoir-to-reservoir variability when investigating the link between connectivity, dynamics, and performance. As physical reservoir computers become more prevalent, developing a systematic approach to network design is crucial. In this article, we examine Random Boolean Networks (RBNs) and demonstrate that specific distribution parameters can lead to diverse dynamics near critical points. We identify distinct dynamical attractors and quantify their statistics, revealing that most reservoirs possess a dominant attractor. We then evaluate performance in two challenging tasks, memorization and prediction, and find that a positive excitatory balance produces a critical point with higher memory performance. In comparison, a negative inhibitory balance delivers another critical point with better prediction performance. Interestingly, we show that the intrinsic attractor dynamics have little influence on performance in either case.
\end{sciabstract}

\section{Introduction}

\sloppy
Reservoir Computing (RC) is a promising field for Machine Learning (ML), as the nonlinear reservoir requires no learning and the readout layer only needs linear regression \cite{Maass2002b,Jaeger2004}, reducing time and computational cost \cite{Schrauwen2007}. Furthermore, it has potential for real-world implementations as physical reservoirs and dedicated Neuromorphic chips do not always possess the ability to adapt \cite{Benjamin2014,Merolla2014,Tanaka2019}. Around the same time in 2002, two models were developed: the Liquid State Machine (LSM) \cite{Maass2002b} and the Echo State Network (ESN) \cite{Jaeger2010} (rectified version). These approaches differ in their neural models, with LSM using time-event-based neurons and ESNs using Artificial Neural Networks (ANN) with continuous activation functions \cite{Jaeger2010}. The Random Boolean Network (RBN) \cite{Bertschinger2004c}, with binary neurons, is a particularly promising model for LSM and allows for a direct relationship between the reservoir design and its performance in a task \cite{Bertschinger2004c,Natschlager2005,Snyder2013}. It is widely used to model and implement reservoirs \cite{Rosin2015,Burkow2016,Echlin2018,Komkov2021a}. 

Studies on the RBN have demonstrated the existence of a phase transition in the dynamics of the reservoir for specific connectivity parameters. Close to the critical regime, an increase in performance in solving various tasks has been reported  (boolean logic operations \cite{Bertschinger2004c}, bit-parity check \cite{Bertschinger2004c}, prediction of Mackey-Glass time series \cite{Canaday2018}). As of now, most studies in the field of RC rely on phase diagrams to exhibit a statistical relationship between connectivity, dynamics, and performance \cite{Bertschinger2004c}, \cite{Busing2010}, \cite{Snyder2013}, \cite{Krauss2019a}, \cite{Metzner2022}. These results have been obtained by considering a limited number of reservoirs (from one \cite{Metzner2022}, to 10 \break \cite{Bertschinger2004c}, up to 100 \cite{Krauss2019}), and with a limited resolution in terms of the control parameter, due to the computational cost of these phase diagrams.

While phase diagrams are essential to comprehend the full range of the computational capabilities these systems can offer, one crucial point is rarely discussed. Since the reservoirs are randomly generated, there might be huge differences between them even though the statistics of their connectivity are the same. Indeed, close to the critical point, reservoir steady-state activities exhibit a wide range of dynamics as discussed by statistical studies \cite{Kinouchi2006}, \cite{DelPapa2017a}, \cite{Krauss2019}, and attractor classification \cite{Rieffel2014}, \cite{Bianchi2016}, \cite{Krauss2019}, \cite{Krauss2019a}, \cite{Metzner2022}.

This article aims at studying the variability of reservoir dynamics, performance, and their correlation. We consider randomly generated RBNs with a single control parameter related to the inhibitory/excitatory balance \cite{Krauss2019a}, tuned with high resolution to perform reliable statistical analysis. We study the excitatory/inhibitory balance, attractor dynamics, and performance, and show that the relationship between the three is more complex than previously thought. In line with the work of \cite{Metzner2022} on ESN, our research reveals that the RBN also possesses two critical points. Depending on whether the balance is in the majority excitatory or inhibitory, we show that reservoirs respectively exhibit optimal performance in either memory or prediction. 

The article is organized as follows: in section \ref{met:main}, we describe the model and prove that it is controlled by the ratio of the standard deviation and mean of the weight distribution (noted $\sigma^{\star}$), which we use to perform all subsequent analyses. In section \ref{p2:activity}, we show that the sign of $\sigma^{\star}$ produces two critical regimes. In section \ref{p2:dominant}, we classify the activity of free-running reservoirs into four classes according to their attractor dynamics for these two critical regimes. We show that each reservoir can be associated with its most dominant attractor. In section \ref{p3:performance}, we evaluate the relationship between connectivity, dominant attractor, and performance in memory and prediction tasks. We then investigate the relationship between the performances of the two tasks, critical regimes, and dominant attractors in section \ref{p3:cross}. This allows us to derive specific recommendations for simplifying the random generation process of reservoirs. Finally, we discuss our findings in section~\ref{conclusion} and their implications for future works in section~\ref{future}. 

\section{Model} \label{met:main}

\begin{figure*}
    \centering
    \includegraphics[scale=0.4]{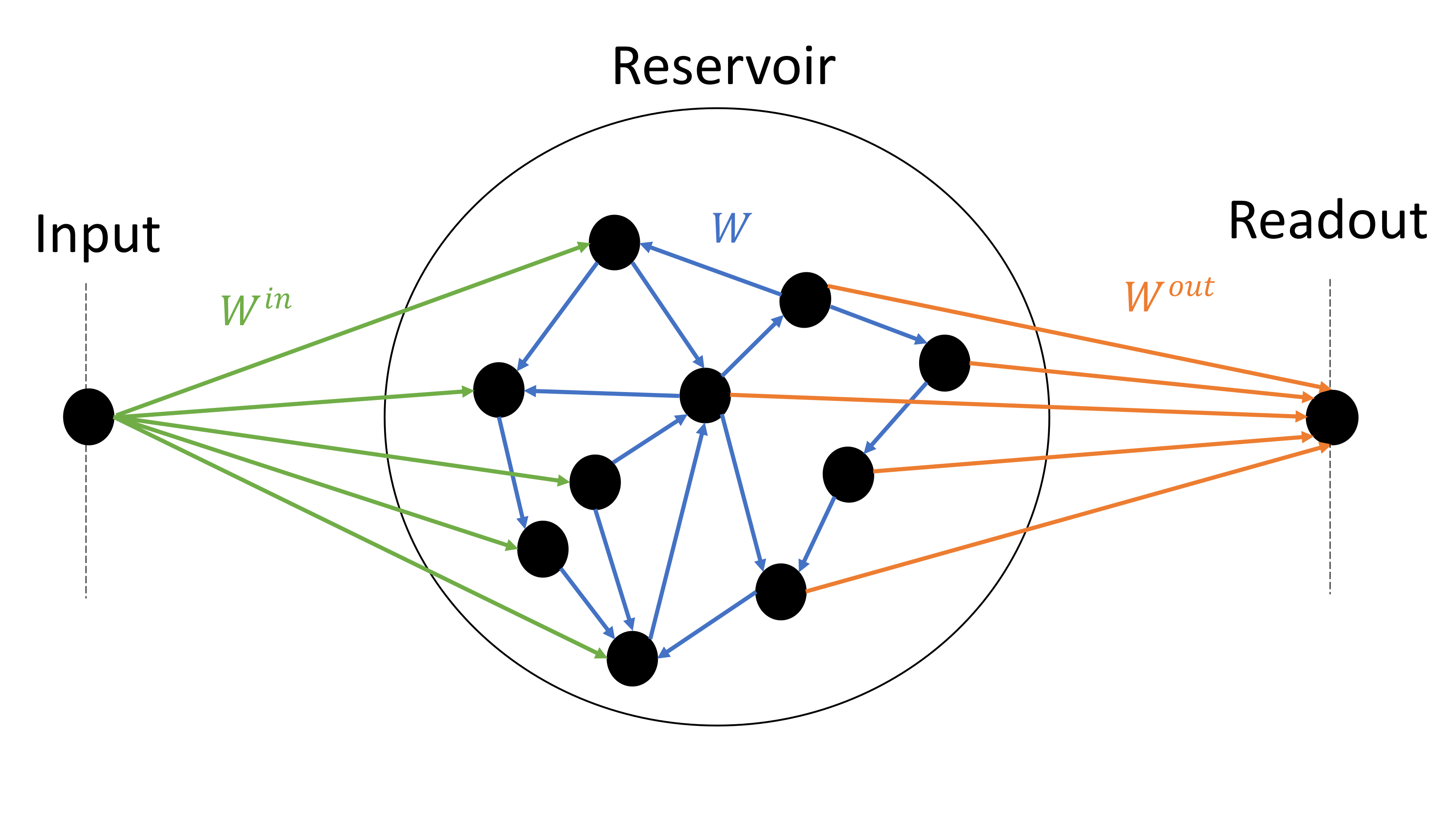}
    \caption{Schematics of the model. The input node (left) randomly projects synaptic weights to half of the reservoir (center) (\textcolor{midgreen}{green}); the reservoir is composed of random recurrent connections (\textcolor{midblue}{blue}); the readout (right) receives input from the other half of the reservoir (\textcolor{darkorange}{orange}).}
    \label{fig:res}
\end{figure*}

The model consists of one input node, the reservoir itself, and an output node (Fig.~\ref{fig:res}). Half of the neurons inside the reservoirs are connected to the input, and the other half to the readout. Thus information between the input and the readout has to pass through the reservoir. The following subsections describe each component and how they are interconnected.

\subsection{The reservoir}

Phase transitions occur \textit{stricto sensu} only in infinite systems, and critical phenomena are easier to observe in large systems \cite{Lavis2021}. As such, we use an RBN model of size $N=10,000$ neurons, which is considerable compared to similar studies in the literature  \cite{Natschlager2005, Busing2010, Metzner2022}. The binary state $x_i(t)\in \{0,1\}$ of the neuron $i$ at the time-step $t$ (with $t \in \mathbb{N}$), is given by: 
\begin{equation}
    x_i(t)=  \theta \left(u_i(t) + \sum_{j=1}^N \textcolor{blue}{w_{ij}}x_j(t-1) \right)
    \label{eq:1}
\end{equation}
\noindent where $\theta$ is the Heaviside step function: $\theta(x)=1$ if $x>0$ and $\theta(x)=0$ otherwise. Each neuron receives the same number of non-zero connections $K=16$, in the range of values shown to display sharp phase transitions \cite{Busing2010}. The non-zero recurrent weights of the reservoir $w_{ij}$ (blue arrows in Fig.~\ref{fig:res}) are i.i.d. and drawn from the Normal or Gaussian density probability function $\mathcal{N}(\mu, \sigma)$. $u_i(t)$ is the external input of the neuron $i$ at times $t$.

\subsection{Input node} \label{met:input}

The input layer reduces to one node, receiving the time series $u(t)$. The input at times $t$ of a given neuron $i$ is:
\begin{equation}
    u_i(t) = \textcolor{classicgreen}{w^{in}_i} u(t)
    \label{eq:2}
\end{equation}
 Where the input weight $w^{in}_i$, of neuron $i$ (green arrows in Fig.~\ref{fig:res}) is drawn from a uniform distribution within $[-0.5, 0.5]$, and half of the weights are set to zero. According to Eq. \ref{eq:1}, if the amplitude of the input far exceeds the total contribution of the recurrent weights, then the input mostly controls the dynamics. Our choice of parameters corresponds to an input of zero average and $\sim0.14$ the standard deviation, which is rather low compared to the recurrent weights. We show in part \ref{p3:performance} that this choice makes the dynamics mostly controlled by the recurrent weights, which is the intended behaviour.  

\subsection{Readout} \label{met:learning}

The adaptation mechanism is in the output layer only, which reduces here to one linear node with a sigmoid activation function $f(x)=\frac{1}{1+e^{-x}}$. As such, the output of the network is given by:
\begin{equation}
    y(t) = f(\textcolor{classicorange}{W^{out}} \vec{x} + c)
    \label{eq:3}
\end{equation}
\noindent Since all experiments consist in reproducing a unidimensional time series, the output $y$ is a scalar as well. The column vector $\vec{x}$ represents the state of the reservoir neurons, while the output weights $W^{out}$ (orange arrows in Fig.~\ref{fig:res}) are stored in a row vector of size $N$, with half of them set to zero. Lastly, the scalar $c$ is the bias. The training is performed with a mean square error (MSE) loss function. Since we had a focus on collecting high-quality data regarding the link between connectivity and performance, we chose the ADAM optimizer \cite{Kingma2015} over the more standard Ridge regression \cite{Burkow2016} often used in the literature. The implementation is made with the PyTorch library, and parameters $\alpha=0.001$, and $4000$ epochs (See Supplementary Materials \ref{suppm:training} for more details). 

\subsection{Connectivity: the control parameter $\sigma^{\star}$} \label{p1:control_param}

To study the reservoir dynamics, one needs the proper definition of a control parameter. Previous work on the RBN often focuses on the average and variance of the recurrent weight matrix (\cite{Bertschinger2004c}, \cite{Natschlager2005}). In the following, we demonstrate the existence of only one control parameter $\sigma^\star$ defined by:

\begin{equation}
    \sigma^{\star}=\sigma/\mu, \quad \mu \neq 0
    \label{eq:4}
\end{equation}

\noindent Where $\mu$ is the mean of the weights and $\sigma$ their standard deviation. Here we study the reservoir in the absence of external excitation, $u_i(t)=0$ in Eq.~(\ref{eq:1}). Let us consider two reservoirs with the same architecture, the same initial state, and with respective weights matrices $W$ and $\lambda W$ with the scalar $\lambda>0$. Since $\lambda>0$, then $\theta(\lambda x)=\theta(x)$, $\forall x$. Thus, according to Eq.~(\ref{eq:1}) for $u_i(t)=0$, the two networks are always in the same state. Thus $(\lambda\mu,\lambda\sigma)$ gives rise to the same time evolution as $(\mu,\sigma)$. The two corresponding reservoirs are totally equivalent. We face two cases depending on $\mu$:
\begin{itemize}
    \item When $\mu=0$, choosing $\lambda=1/\sigma$ leads to the conclusion that all reservoirs $(0, \sigma)$ are strictly equivalent to the reservoir $(0,1)$. Hence reservoirs with $\mu=0$ are independent of $\sigma$.
    \item When $\mu\neq0$, choosing $\lambda=1/|\mu|$ leads to the weights of the second reservoir distributed with a mean of $\pm1$ and a standard deviation $\sigma/|\mu|$. Hence we define the control parameter of the RBN as in Eq.~(\ref{eq:4}).
\end{itemize}

Eq.~(\ref{eq:4}) characterizes the distribution of the weights: the mean is the sign of $\sigma^\star$, and the standard deviation is its absolute value. Other distribution characterizations directly relate to $\sigma^{\star}$. For instance, it is controlling the balance $b$ between excitation and inhibition, defined by \cite{Krauss2019a} as:

\begin{align}
    b=(S_+ - S_-)/S \\
    S_\pm = \frac{S}{2}(1\pm b)
    \label{eq:5}
\end{align}

With $S=S_+ + S_-=KN$ the total number of synapses, $S_-$ the number of inhibitory synapses ($w_{ij}<0$), and $S_+$ the number of excitatory synapses ($w_{ij}>0$). By taking a normal weight distribution, the number of excitatory synapses is given by: 

\begin{equation}
    S_+ = S \int_0^{+\infty} \frac{1}{\sqrt{\pi}}  e^{-\left( x-\frac{\mu}{\sqrt{2}\sigma} \right)^2}dx
    \label{eq:6}
\end{equation}

\begin{figure*}
    \centering
    \includegraphics[scale=0.6]{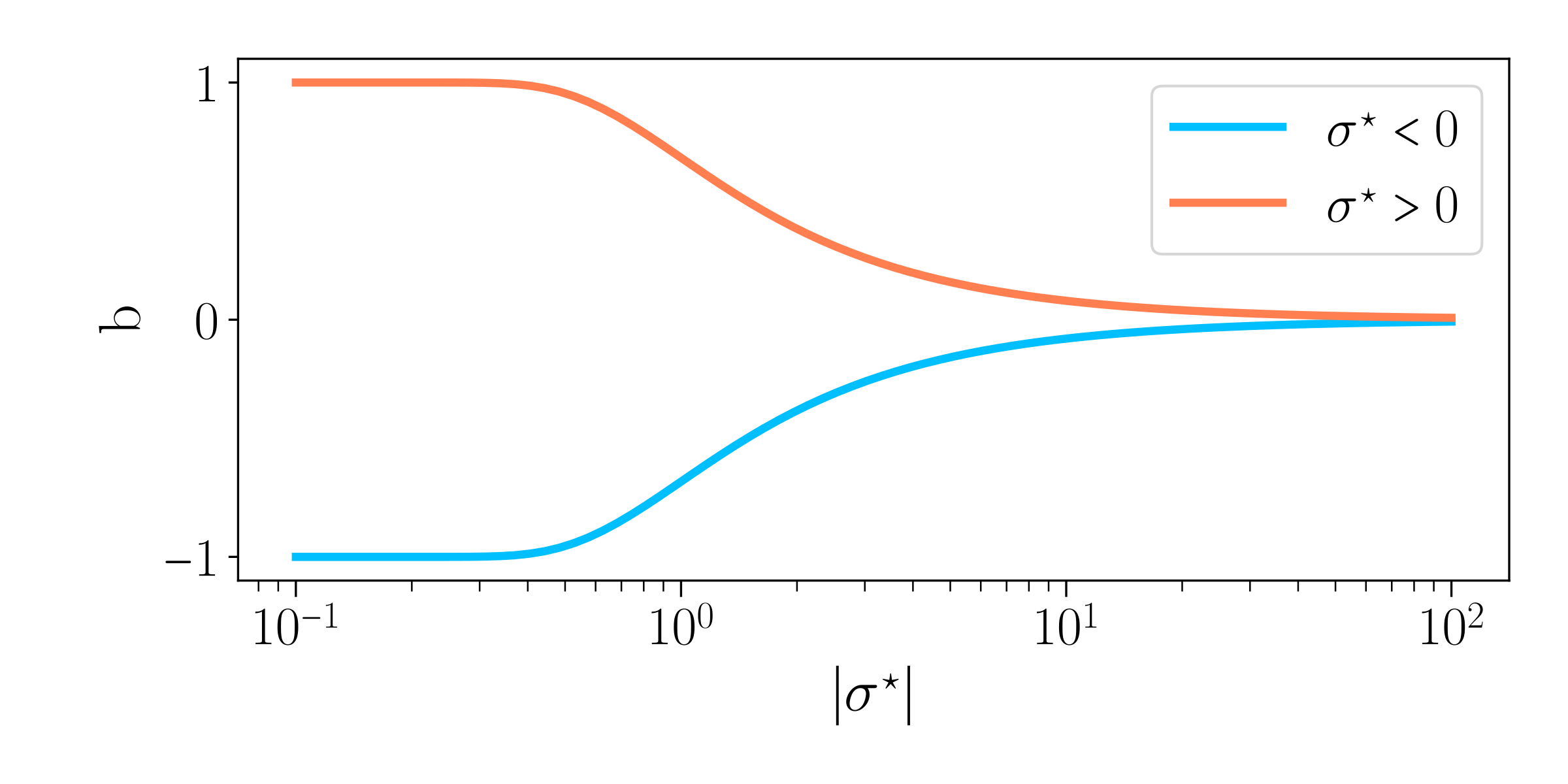}
    \caption{Excitation/inhibition balance $b$ as a function of the absolute value of the connectivity parameter $\sigma^\star$, as defined in Eq.~\ref{eq:4}, for $\sigma^\star<0$ (\textcolor{cyan}{\rule[0.5ex]{1em}{0.6pt}}) and $\sigma^\star>0$ (\textcolor{darkorange}{\rule[0.5ex]{1em}{0.6pt}}). When the average of weights is positive ($\sigma^\star>0$), whereas the reverse is true when the average of weights is negative.}
\label{fig:balance}
\end{figure*}

By substituting Eq.~(\ref{eq:6}) in Eq.~(\ref{eq:5}), we find $b=\mathrm{Erf}[1/(\sqrt2\sigma^\star)]$, with $\mathrm{Erf}$ the error function. Thus, by controlling the weight distribution, our control parameter $\sigma^\star$ drives the excitatory to inhibitory balance and thus the reservoir dynamics, in line with \cite{Krauss2019} and \cite{Metzner2022}. Fig.~\ref{fig:balance} shows the relationship between $b$ and $\sigma^\star$. The case $\mu=0$ corresponds to $b=0$ (perfect balance between excitation and inhibition) and $\sigma^\star\to\infty$, for any value of $\sigma$. For $\sigma^\star < 0$, $b \in [-1, 0]$, i.e. there is a majority of inhibitory synapses while for $\sigma^\star > 0$, $b \in [0, 1]$, hence a majority of excitatory synapses. 

To finish, the spectral radius $\rho(W)$ of the weight matrix $W$ is a particularly relevant quantity in the context of the ESN, a continuous version of reservoirs, where the critical point corresponds to a spectral radius of one. However, in the case of discontinuous activation functions, such as the one we have with the RBN, it has been shown that the Echo State Property (ESP) cannot be achieved. The spectral radius alone fails to characterize the dynamics and performance of these reservoirs \cite{Oztuik2007}, \cite{Alexandre2009}, \cite{Tieck2018}, \cite{Balafrej2022}. In supplementary material \ref{suppm:rho}, we explicitly discuss the link between $\rho$ and the mean and variance of the weight matrix and show $\rho$ is of no particular interest in the study of the dynamics.

As a consequence, in the following, we will use $\sigma^{\star}$ as the unique control parameter (in the range of values displayed in supplementary material \textbf{S}\ref{suppm:sigma_values}). 

\section{Statistics of dynamics at the critical points} \label{p2:main}

The importance of neural networks dynamics in understanding their performances has been widely explored (\cite{Bertschinger2004c}, \cite{Busing2010}, \cite{Krauss2019a}, \cite{Metzner2022}). The purpose of this part is to investigate the relationship between connectivity and dynamics through two statistical analyses: 
\begin{enumerate}
    \item \textbf{Activity statistics:} In section \ref{p2:activity}, we analyze the statistics of the neural activity as a function of the control parameter $\sigma^{\star}$. We demonstrate the existence of two critical points and characterize them.
    \item \textbf{Reservoirs attractors:} In section \ref{p2:dominant}, we classify the steady state time evolution of the network activity into four distinct attractors and study the influence of the initial state and random weight generation. We show that reservoirs possess a dominant attractor independent of initial conditions.
\end{enumerate}

\subsection{Statistics of the activity of free-evolving reservoirs} \label{p2:activity}

\subsubsection{Methodology:}
The first experiment is a free evolution of reservoirs in the absence of input, i.e. $u_i=0$. We define the network activity as $A(t)=\sum_ix_i(t)/N$, with $N$ the number of neurons in the reservoir, and $A\in [0,1]$. $A$ is also the proportion of excited neurons: $A=0$ if the network is extinguished ($x_i=0 \quad \forall i$), $A=1$ if the network is saturated ($x_i=1 \quad \forall i$). At the initial state, we randomly force $20\%$ of neurons to an up state ($x_i=1$), i.e. $A(t=0)=0.2$. After a transient regime of $1000$ time steps, the reservoir reaches a steady state where we perform statistics. In the following, $A$ will refer to the activity measured in that steady state (see supplementary material \textbf{S}\ref{suppm:measure} for a more formal definition). For each value of $\sigma^{\star}$, we perform statistics on $100$ randomly generated reservoirs (see supplementary material \textbf{S}\ref{suppm:nolearning} for more details on the experiment).

\subsubsection{Analysis:}
In the following, a bar over a variable $\overline{(.)}$ represents an average over time for a given reservoir, while the brackets $\langle . \rangle$ represents an average over different randomly generated reservoirs. In the first analysis, we calculate the time-average steady activity $\bar{A}$ for a given reservoir and its time-variance $\overline{\delta A^2}$, where we define $\delta A=A-\bar A$. We average these quantities over the reservoirs to give $\langle\bar A\rangle$ and $\langle \overline{\delta A^2} \rangle$ for each value of $\sigma^\star$. Next, we evaluate the average and variance over reservoirs of the binary entropy $H_b$, or BiEntropy \cite{Croll2014} of the time-dependent activity. Compared to the Shanon entropy, the advantage of this metric is that it can discriminate ordered from disordered strings of binary digits. It has been used in machine learning \cite{Mamun2016}, \cite{Zhou2022}, but to our knowledge, this is the first time in reservoir computing. The binary entropy varies between 0 for fully ordered bit-streams and 1 for fully disordered ones. We compute the BiEntropy of the binarized time dependence of the steady activity for each reservoir (for the exact definition of all the metrics, see \textbf{S}\ref{suppm:metrics}).

\subsubsection{Results:} 
\begin{figure*}
    \centering
    \includegraphics[scale=0.5]{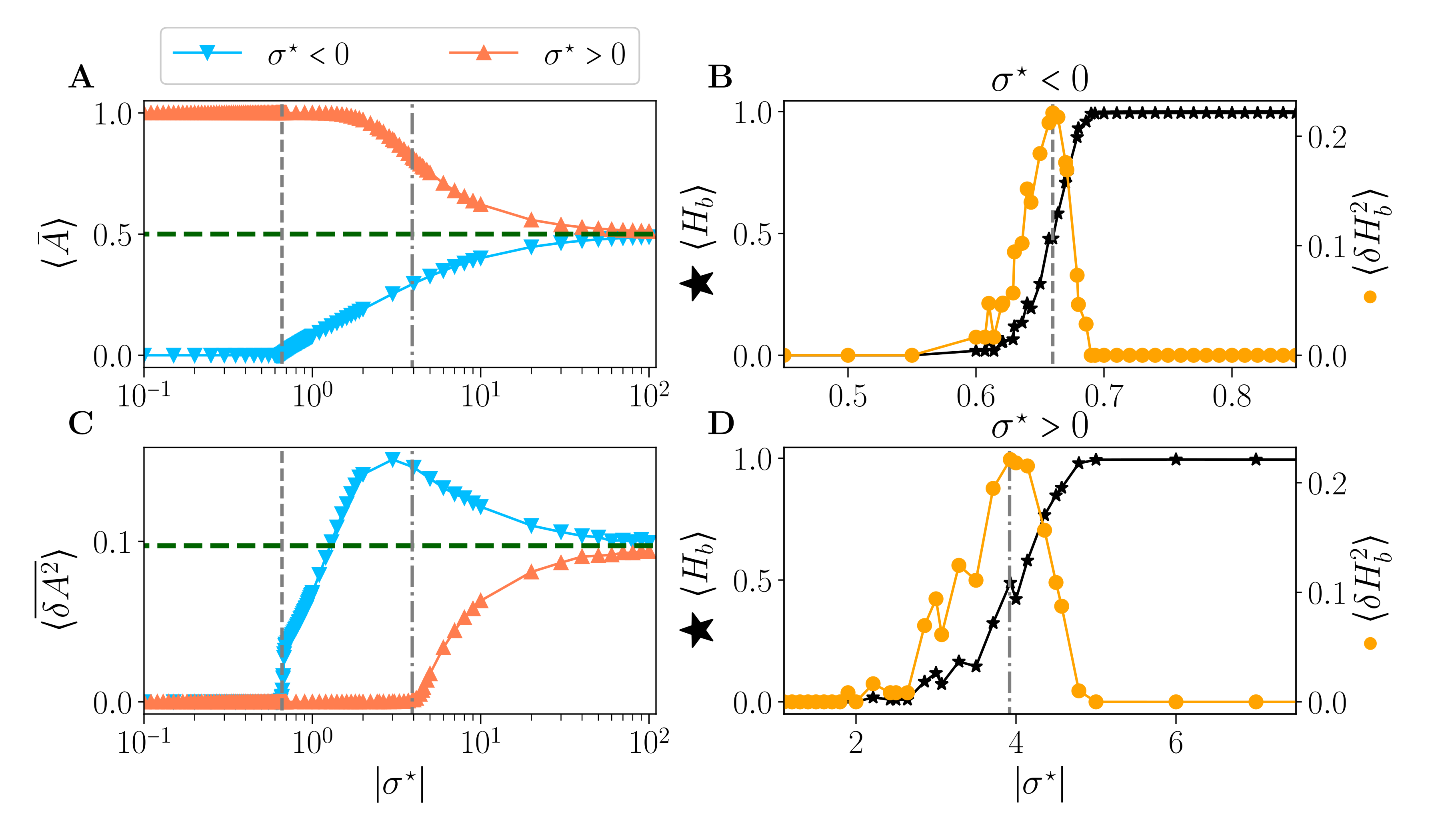}
    \caption{Statistics of the activity of free running reservoirs in the steady state as a function of $|\sigma^{\star}|$. Each dot represents the statistics over $100$ reservoirs ran once. Average over reservoirs of time average activity $\langle\bar A\rangle$ (\textbf{A}), and average over reservoir of time variance $\langle \overline{\delta A^2} \rangle$ (\textbf{C}), for $\sigma^{\star} < 0$ (\textcolor{cyan}{$\blacktriangledown$}) and $\sigma^{\star} > 0$ (\textcolor{darkorange}{$ \blacktriangle$}). In all plots, the gray vertical lines represent the critical values of the control parameter for $\sigma^{\star}_c<0$ (\dashline) and $\sigma^{\star}_c>0$ (\dashdotline{0.3in}). \textbf{B} and \textbf{D} zoom on the region of interest close to the critical points: average over reservoirs of BiEntropy $\langle H_b\rangle$ (\textcolor{black}{$\star$}, left scale) and BiEntropy variance $\langle\delta H_b^2\rangle$ (\textcolor{classicorange}{$\bullet$}, right scale), for $\sigma^{\star}<0$ (\textbf{B}) and $\sigma^{\star}>0$ (\textbf{D}). }
\label{fig:panel1}
\end{figure*}

The time-average activity $\langle\bar A\rangle$ as a function of $\sigma^\star$ is shown in Fig.~\ref{fig:panel1}\textbf{A}, for both signs of $\sigma^{\star}$ (Fig.~\ref{fig:balance}). The green dashed line represents the value obtained for $\mu=0$, i.e. $\sigma\rightarrow \infty$. The perfect balance in excitation ($b=0$) results in half of the neurons being activated $\langle\bar A\rangle=0.5$. The variance $\langle\overline{\delta A^2}\rangle$ vs $\sigma^\star$ is shown in Figs.~\ref{fig:panel1}\textbf{C}. For the lowest values of $|\sigma^{\star}|$, the reservoirs are frozen (zero variance) either extinguished (for $\sigma^\star<0$) or saturated ($\sigma^\star>0$). This corresponds to reservoirs being respectively purely inhibitory ($b=-1$) or excitatory ($b=1$). Already at the level of the statistics of the activity, there is a clear difference between both signs of $\sigma^{\star}$: for $\sigma^\star<0$ there is a threshold in $\sigma^{\star}$ (vertical dashed line at $\sigma^{\star}\sim-0.7$) above which the average activity and its variance rise abruptly and simultaneously. In contrast, for $\sigma^\star>0$, there is a wide region where no dynamic is detected (zero variance), yet the network is not saturated but its activity decays continuously. The variance starts rising at $\sigma^{\star}>4$ (vertical dash-dotted line). 

The average BiEntropy $\langle H_b \rangle$ vs. $|\sigma^{\star}|$ is plotted in Figs.~\ref{fig:panel1}\textbf{B} for $\sigma^{\star}<0$ and ~\ref{fig:panel1}\textbf{D} for $\sigma^{\star}>0$ (left scale, black stars on both plots). These two plots zoom in the vicinity of the phase transition, as statistics are stationary elsewhere. There is a continuous transition between a fully ordered phase ($\langle H_b \rangle=0$) and a fully disordered one ($\langle H_b \rangle=1$). Since the BiEntropy is a measure of order, these results suggest that the transition we observe is related to the apparition of chaos in the reservoir above a critical value of $\sigma^\star$ (\cite{Langton1991}, \cite{Bertschinger2004c}, \cite{Rieffel2014}, \cite{Kusmierz2020}). The variance of the BiEntropy $\langle\delta H_b^2\rangle$ is shown in Figs.~\ref{fig:panel1}\textbf{B} for $\sigma^{\star}<0$ and ~\ref{fig:panel1}\textbf{D} for $\sigma^{\star}>0$ (right scale, orange circles on both plots). It is zero when either in the ordered or disordered phase and spikes at the transition. Its maximum coincides with $\langle H_b \rangle \simeq 0.5$: the variance of BiEntropy captures the edge of chaos as a balance between order and disorder. More striking, it also coincides with the position at which the variance of the activity rises (vertical dashed lines). The peak of $\langle\delta H_b^2\rangle$ thus provides a clear definition of the position of two critical points: $\sigma^\star_c\simeq -0.66$ and $\sigma^\star_c\simeq 4.0$, which correspond respectively to critical balances $b_c\simeq -0.87$ (94\% of inhibitory synapses) and $b_c\simeq 0.19$ (60\% of excitatory synapses). Moreover, the transition between order and disorder is much wider for $\sigma^\star>0$. This asymmetry between both signs of $\sigma^\star$ is a property of our model since $\theta(-x)\neq\pm\theta(x)$ in Eq.(\ref{eq:1}). From now on, we will refer to the \textit{critical points} as the point where the maximum of BiEntropy variance is obtained, and we will define the \textit{critical regions} as the regions with $\langle\delta H_b^2\rangle\neq0$. 

\subsubsection{Discussion:} 
 
Similar to \cite{Krauss2019} and \cite{Metzner2022}, the existence of phases separated by critical points as a parameter is varied is reminiscent of the phase diagrams drawn in thermodynamics. If we associate the state of a neuron, 0 or 1, to the state of an Ising spin, either down or up, then $\langle\bar A\rangle$ corresponds to the average magnetization per spin of the network and $\langle\overline{\delta A^2}\rangle$ to the variance of its fluctuations, i.e. magnetization noise. At equilibrium, it is proportional to the magnetic susceptibility according to the fluctuation-dissipation theorem \cite{Callen1951}. The total magnetization plays the role of an order parameter, and the transition order is obtained by considering discontinuities, as a function of temperature, of the order parameter and its derivatives with respect to the external field \cite{Landau1980}. Here we observe that the average activity is always continuous as a function of $\sigma^\star$. At the same time, $\langle\overline{\delta A^2}\rangle$ is continuous for $\sigma^\star>0$ but shows a discontinuity at the critical point for $\sigma^\star<0$. This strongly suggests that the two "phase transitions" are of a different type.

\subsection{Dominant attractor of reservoirs} \label{p2:dominant}

In the previous section, we considered the \emph{average} behaviour of reservoirs: for a given value of $\sigma^\star$ we averaged over many realizations of the distribution of synaptic weights. However, from a practical point of view, one wants to use one network to work with different inputs. This raises two questions: that of the reservoir-to-reservoir variability (do all reservoirs behave similarly?) and that of the sensitivity of a given reservoir to initial conditions. We address these questions in this section.

\subsubsection{Methodology:}
We submitted our reservoirs again to a free evolution without input ($u_i(t)=0$). For each value of $\sigma^{\star}$, we created $100$ reservoirs with randomly tossed weight matrices. Each reservoir is run $100$ times, with a different random initial state, of activity $A(t=0)=0.2$ (for more details, see \textit{Statistics of reservoirs} in \textbf{S}\ref{suppm:experiment}). 

\subsubsection{Analysis:} 
We classify the attractor obtained in the steady-state activities, as proposed in \cite{Krauss2019}. We categorize the activity signals into one of the four types of attractors (see \textbf{S}\ref{suppm:classification} for a grounded justification of each category):
\begin{itemize}
    \item \textit{Extinguished} activity: The steady-state activity $A(t)$ is always zero. This means that the initial activity died during the transient phase and that the reservoir could not propagate it further in time. For simplicity, we will sometime refer to it as \textit{dead} attractor.
    \item \textit{Fixed} point attractor: The steady-state reservoir is active ($A(t)\neq0$), but the activity is independent of time ($\delta A(t)=0$). This includes the \textit{saturated} states $A(t)=1$ of ref. \cite{Rieffel2014}. 
    \item \textit{Cyclic} attractor: $A(t)$ is periodic with a periodicity larger than one time-step.
    \item \textit{Irregular} attractor: $A(t)$ is neither constant nor periodic within the duration of the simulation. Note that since the RBN is finite, discrete, and deterministic, given enough time, any sequence of states should eventually repeat, taking at most $2^N$ time steps.
\end{itemize}
We determine the attractor obtained at the steady state for each reservoir and initial condition. We then compute the distribution of attractors for each value of $\sigma^\star$ obtained overall the initial conditions of all reservoirs. The statistics are thus computed on 10,000 steady activities for each $\sigma^{\star}$.

\begin{figure*}
\centering
    \includegraphics[scale=0.6]{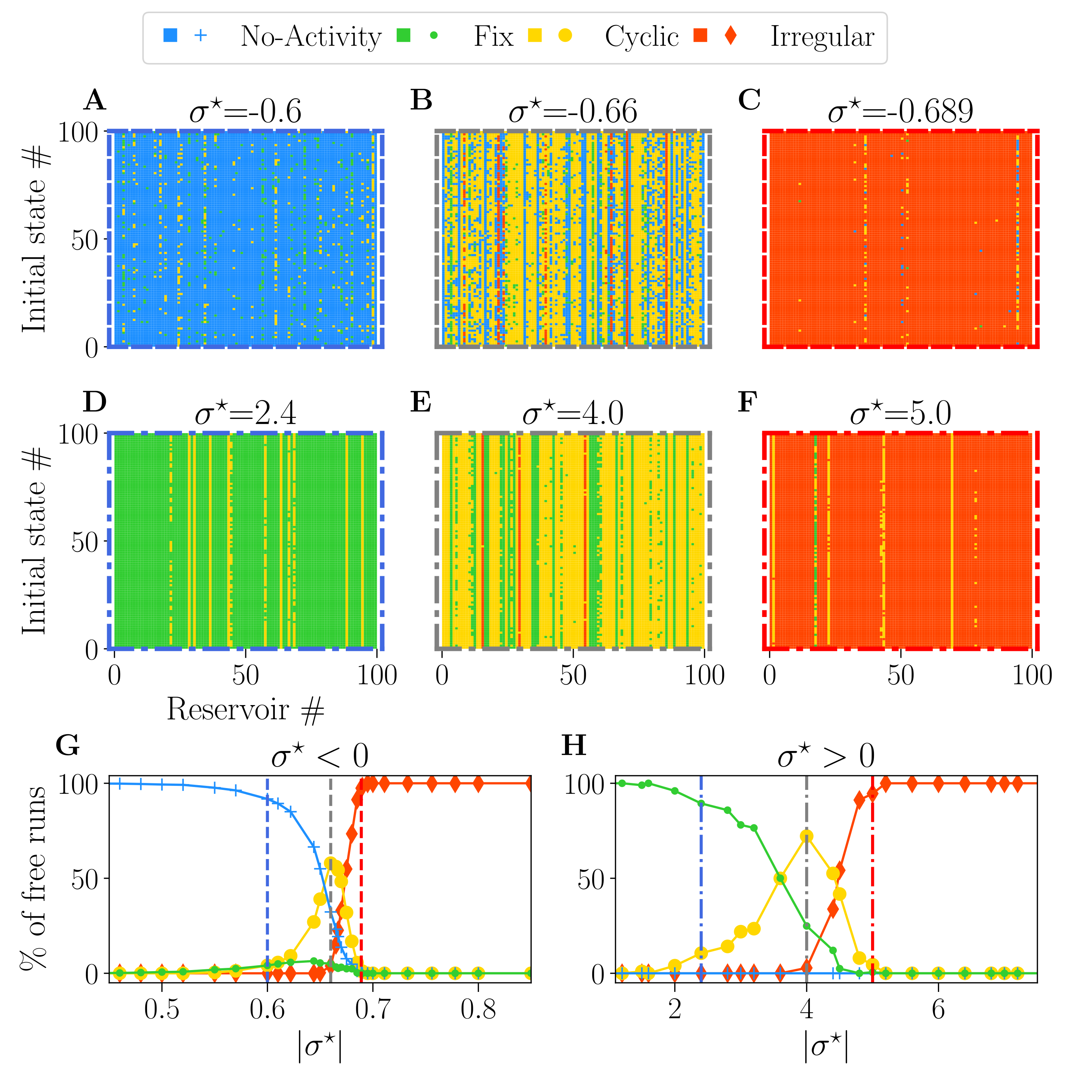}
    \caption{The attractor landscape of reservoirs: for $\sigma^{\star}<0$ (\textbf{A}, \textbf{B}, \textbf{C}, \textbf{G}), and for $\sigma^{\star}>0$ (\textbf{D}, \textbf{E}, \textbf{F}, \textbf{H}). The influence of initial conditions for specific values of $\sigma^{\star}$: (\textbf{A}) $\sigma^{\star}=-0.6$, (\textbf{B}) $\sigma^{\star}=-0.66$, (\textbf{C}) $\sigma^{\star}=-0.689$, (\textbf{D}) $\sigma^{\star}=2.4$, (\textbf{E}) $\sigma^{\star}=4.0$, (\textbf{F}) $\sigma^{\star}=5.0$. In each plot, the vertical axis represents different numbers (\#) of initial random states of free-running reservoirs. The horizontal axis represents different numbers (\#) of reservoirs with various initial weight tossing, randomly generated with distinct seeds. Pixels of colours represent the attractor obtained at the steady state, with the same colours as (\textbf{G}, \textbf{H}). \textbf{G}, \textbf{H}: percentage of steady-state activities belonging to each category of attractors: no-activity (\textcolor{dead}{$+$}), fix (\textcolor{fix}{$\sbullet$}), cyclic (\textcolor{cyclic}{$\bullet$}), and irregular (\textcolor{chaos}{$\blacklozenge$}). Each dot represents the statistics over $100$ reservoirs ran $100$ times, hence $1000$ runs. Each (\textbf{G}) $\sigma^{\star}<0$, and  $\sigma^{\star}>0$ . In each row (\textbf{A}, \textbf{B}, \textbf{C}) and (\textbf{E}, \textbf{F}, \textbf{G}), the coloured dashed boxes surrounding the plots correspond to the values of $\sigma^{\star}$, indicated as vertical lines in plots \textbf{G} and \textbf{H}.}
\label{fig:panel2}
\end{figure*}

\subsubsection{Results:}
Fig.~\ref{fig:panel2}\textbf{A-F} provide examples of attractors, encoded in the colours, for each reservoir (x-axis) and each initial condition (y-axis) for different values of $\sigma^\star$. The left column (blue-bordered boxes) corresponds to values below the critical point (vertical blue lines on Figs.~\ref{fig:panel2}\textbf{G,H}), the center column (gray-bordered boxes) to values at the critical point (vertical gray lines), and the right column (red-bordered boxes) to values above the critical points (vertical red lines). The upper row displays negative $\sigma^\star$ values, while the lower row features positive $\sigma^\star$ values.

Away from the critical point, a dominant colour is observed, meaning that reservoirs exhibit a dominant attractor. Steady activities are predominantly extinguished and fixed on the left side of the critical point (\textbf{A}, \textbf{D}) and irregular at the right (\textbf{C}, \textbf{F}). Close to the critical points (\textbf{B} and \textbf{E}), there is an increase in the diversity of attractors, as previously observed  \cite{Karimipanah2017}.

 Fig. \ref{fig:panel2}.\textbf{G} and \textbf{H} show the statistical distribution of all obtained attractors versus $|\sigma^\star|$. As expected from the previous analysis, there is no attractor diversity on the far left and right of the plots, as we obtain one primary attractor. Dead (blue line) or fixed (green line) attractors are found for low values of $|\sigma^\star|$, and their proportion decays slowly across the transition. Within the critical region coexist all attractors in various proportions. Chaotic attractors start to appear precisely at the transition (vertical gray lines), while the domain where cyclic attractors exist coincides with the critical region of nonzero BiEntropy variance (Fig.~\ref{fig:panel1}.\textbf{B,D}). The point at which cyclic attractors are most present is also precisely $\sigma^\star_c$. These results corroborate what we inferred in the previous section: on the disordered phase $|\sigma^\star|>|\sigma^\star_c|$, attractors are irregular, while the ordered phase is characterized by fixed or dead attractors. We note an asymmetry between both sides of the transition: irregular attractors appear only in the disordered phase. From the point of view of the attractors, both signs of $\sigma^\star$ lead to similar behaviours, except again, that the transition region is much wider for $\sigma^\star>0$.

\subsubsection{Discussion:}
\begin{figure*}
    \centering
    \includegraphics[scale=0.5]{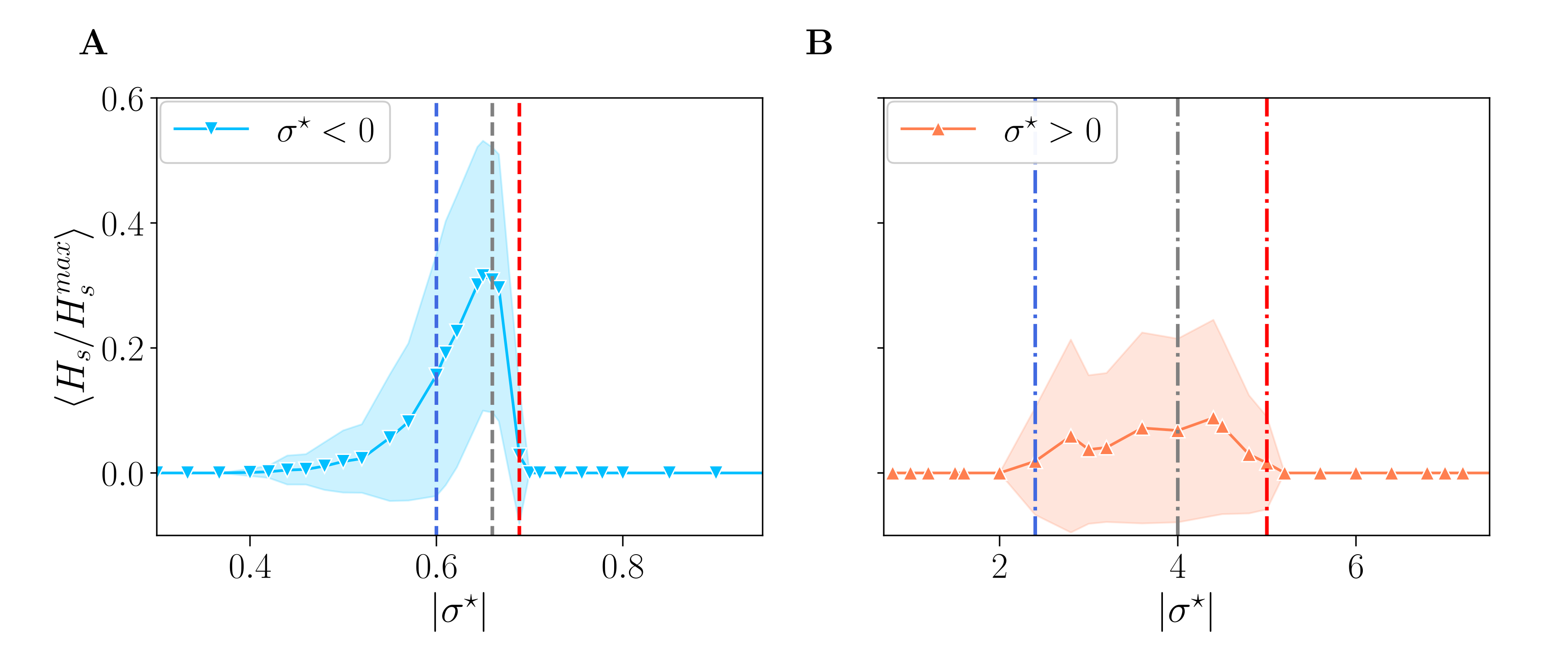}
    \caption{The normalized Shanon entropy of individual reservoir attractor distributions. \textbf{A} and \textbf{B}: The average over reservoirs entropy of the entropy $H_s$ versus the control parameter $|\sigma^\star|$, for $\sigma^\star<0$ (\textbf{A}) and $\sigma^\star>0$ (\textbf{B}). The shaded area represents one standard deviation. The coloured hashed lines correspond to the boxes of Fig.~\ref{fig:panel2}: (blue) $\sigma^\star=-0.6$ (\textbf{A}) and $\sigma^\star=2.4$ (\textbf{D}), (gray) $\sigma^\star=-0.66$ (\textbf{B}) and $\sigma^\star=4.0$ (\textbf{E}), (red) $\sigma^\star=-0.689$ (\textbf{C}) and $\sigma^\star=5.0$ (\textbf{F}). }
\label{fig:panel2s1}
\end{figure*}

Our results suggest that the critical points enhance sensitivity to the initial states and configuration of the weights, explaining the reservoir-to-reservoir variance and increase in dynamic diversity. Reservoirs around the negative $\sigma_c^{\star}$ (Fig.~\ref{fig:panel2}.\textbf{B}) possessed a distribution of attractors with far more variety than the one with positive $\sigma_c^{\star}$ (Fig.~\ref{fig:panel2}.\textbf{E}), further reinforcing the idea that the sign of $\sigma^{\star}$ produces two distinct types of critical regimes. We quantified this in the supplementary material \textbf{S}\ref{suppm:entropy} by computing the entropy of reservoir attractor distributions plotted in Fig.~\ref{fig:panel2s1}. We interpret that result by suggesting that inhibition might be a key factor for enhancing dynamic diversity. 

For the purpose of reservoir design, our findings suggest that with both critical points, most reservoirs possess an attractor obtained predominantly in most trials, independent of the initial state. The statistics of dominant reservoir attractors are presented in supplementary Fig.~\ref{fig:panel2s1}, and found to be similar to the one in Fig.~\ref{fig:panel1}.
The presence of vertical colour lines in Fig.~\ref{fig:panel2}\textbf{A-F} means that, in most cases, the behaviour of the reservoirs does not depend on the initial state, even in the critical region (this is more thoroughly shown in the supplementary material \textbf{S}\ref{suppm:entropy}). As a consequence, a dominant attractor can be associated with each reservoir, irrespective of the initial condition.

\section{What drives performances} \label{p3:main}

In this section, we examine whether there is a relationship between reservoir dynamics in the absence of input, as explored in the previous section, and its ability to perform two demanding tasks: memory and prediction. This is done in two steps:
\begin{itemize}
    \item \textbf{Connectivity, attractor, and performance:} In section \ref{p3:performance}, we analyze the performance obtained separately in each task, depending on the control parameter, for each dominant attractor category. We show that the key factor driving performance is the excitatory/inhibitory balance.
    \item \textbf{Attractor and cross-task performance:} In section \ref{p3:cross}, we analyze all reservoir performances independently of the control parameter. For each reservoir, we study the relationship between the performance obtained in each task and their dominant attractor. This allows us to deduce how to generate a reservoir for the best general purpose.
\end{itemize}

From now on, and for ease of notation, a reservoir with a dominant attractor obtained during free evolution (defined in previous section \ref{p2:dominant}) will be referred to as either: a \textit{extinguished}, \textit{fix}, \textit{cyclic}, or \textit{irregular} reservoir (e.g., an \textit{extinguished reservoir} refers to a reservoir with an extinguished dominant attractor).

\subsection{Performance in memory and prediction tasks} \label{p3:performance}

Close to the critical points, we obtained various dominant attractors for a single value of $\sigma^\star$. This raises an important question regarding the relationship between the dominant attractor of a reservoir and its performance. Specifically, it is worth investigating whether the dominant attractor influences the reservoir's performance. If this is the case, grouping attractor categories by discrete performance levels may be possible based on a single value of $\sigma^\star$.

\subsubsection{Methodolody:} We evaluate the performance of the networks to execute two fundamental tasks: \textit{memory} and \textit{prediction}. Each reservoir receives an input $u(t)$, and the readout target is $T(t)=u(t+\delta)$, equal to the input shifted in time by  $\delta$ time steps. $\delta<0$ corresponds to a memory task, and $\delta>0$ to a prediction task. For each value of $\sigma^{\star}$, we use 100 reservoirs, and each reservoir is run five times, with a different random tossing of the input weight matrix (more detail on the training procedure in \textbf{S}\ref{suppm:training}). 

\begin{figure*}
\centering
    \includegraphics[scale=0.56]{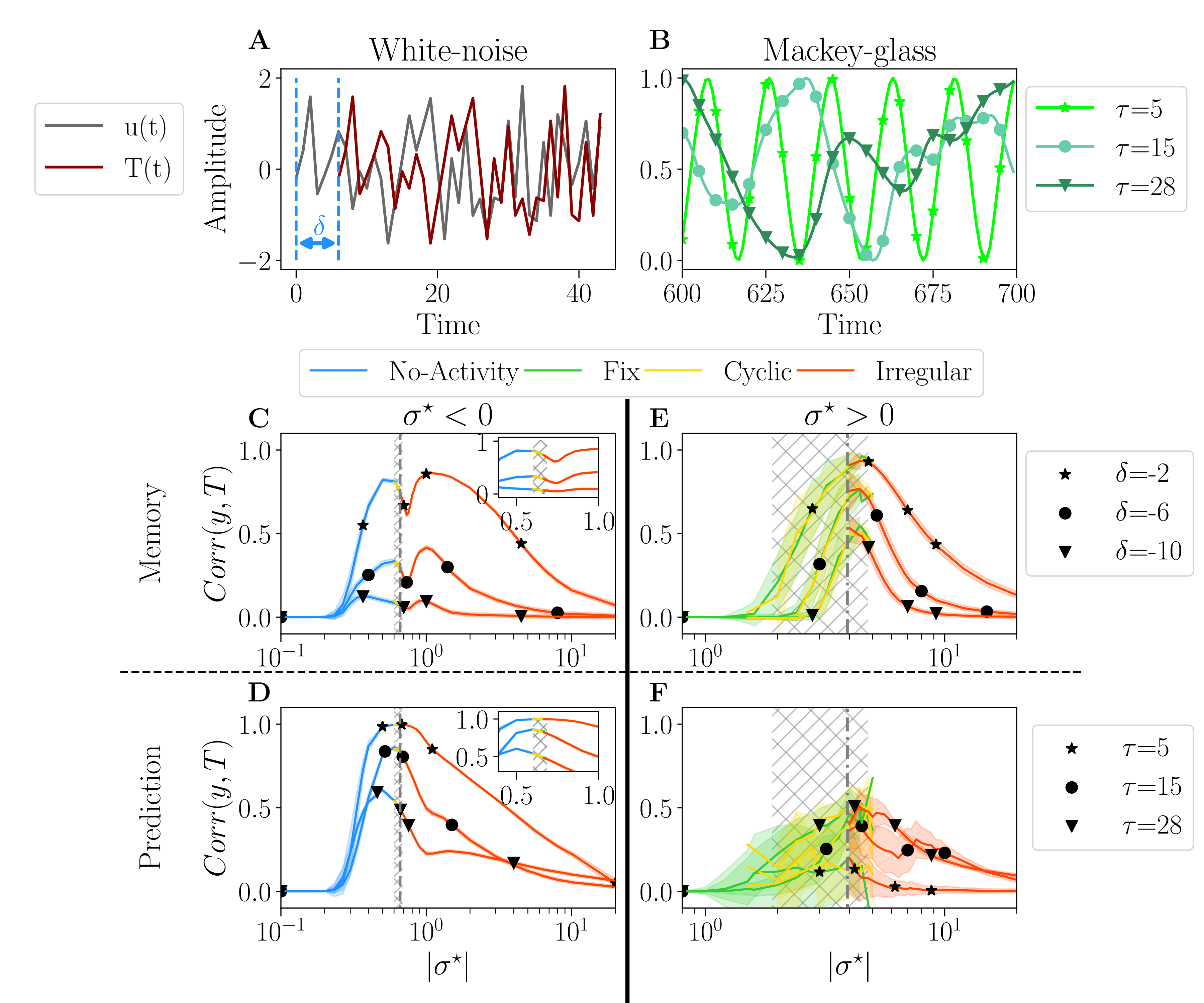}
    \caption{Performances for two tasks: white noise memory (\textbf{C}, \textbf{E}); and Mackey-glass prediction (\textbf{D}, \textbf{F}). \textbf{A} and \textbf{D}: examples of signals for each task with their respective parameters. \textbf{A}: White noise memory task, which consists in remembering the input (\textcolor{darkgray}{gray}), to reproduce it in output (\textcolor{darkred}{dark red}) with a negative delay $\delta$ (shown example corresponds to $\delta=-6$). \textbf{B}: Mackey glass is controlled by the parameter $\tau$ (see methodology \ref{suppm:performance} for more details), ranging from periodic to chaotic. \textbf{C}, \textbf{D}, \textbf{E}, \textbf{F}: the average performance $Corr(y, T)$ between the output $y$ and target $T$, plotted over $|\sigma^\star|$, for each dominant attractor category: \textit{no-activity}, \textit{fix}, \textit{cyclic} or \textit{irregular}. For each value of $\sigma^\star$ we have $100$ reservoirs. The solid line then represents the average over reservoirs belonging to the same attractor category; individual reservoir performances are averaged over $5$ initial conditions. The shaded area represents one standard deviation. Higher correlations indicate better performance. The hatched gray area represents the critical regions, as defined in section \ref{p2:activity}. \textbf{C} and \textbf{D}: the performance in the white-noise memory task; three values of $\delta$ are tested: $-2$ ($\star$), $-6$ ($\bullet$), $-10$ ($\blacktriangledown$). \textbf{E} and \textbf{F}: the performance of Mackey-Glass prediction ($\delta=+10$); three values of $\tau$ are tested: $5$ ($\star$), $20$ ($\star$), $50$ ($\blacktriangledown$). \textbf{C} and \textbf{D}: Performance for $\sigma^{\star}<0$, with inside each plot a zoom on the critical region. \textbf{E} and \textbf{F}: performance for $\sigma^{\star}>0$.}
\label{fig:panel3}
\end{figure*}

The first task consists of memorizing a purely random signal (i.e., uncorrelated white noise), and since there is absolutely no correlation in the input, only memorization is involved. Fig.~\ref{fig:panel3}.\textbf{A} illustrates this task for one value of $\delta=-6$, with white noise as input $u$, and the target $T$. 

For the second task, we explore the ability of the reservoir to predict a time series, $\delta=10$ time steps in the future. The input is the well-known Mackey-Glass time series, as it is a common benchmark of this type of task (\cite{Hajnal2006}, \cite{Goudarzi2016a}, \cite{Canaday2018}, among others), notably testing the ability to infer  non-linear dynamics. The signal regularity is controlled by the parameter $\tau$, see Fig.~\ref{fig:panel3}.\textbf{B}, ranging from periodic with $\tau=5$, to chaotic for $\tau=28$ (more information on the experiments in \textbf{S}\ref{suppm:performance}).

\subsubsection{Analysis:}
The performance of a reservoir is measured by computing the correlation product $Corr(y,T)$ between the output $y$ and the target $T$. A perfect match corresponds to a correlation of one, while a random output gives a zero correlation. An individual reservoir performance score is then obtained by averaging over the initial conditions. Each individual reservoir is associated with its dominant attractor, and the statistics of the performance of reservoirs are performed separately for each attractor.

\subsubsection{Results:} 
The average performance is plotted as a function of $|\sigma^{\star}|$ in Fig.~\ref{fig:panel3}\textbf{C,E} for the memory task and in Fig.~\ref{fig:panel3}\textbf{D,F} for the prediction task. The left column (\textbf{C} and \textbf{D}) corresponds to $\sigma^\star>0$, and the right column (\textbf{E} and \textbf{F}) to $\sigma<0$. The colour of the lines corresponds to the attractor. 

For $\sigma^\star < 0$ (Fig.~\ref{fig:panel3}.\textbf{C} and \textbf{D}), performance increases over a very wide range of $\sigma^\star$, both for memory and prediction. This range includes the critical region (gray hatched area) but is vastly broader. Thus, being within the critical region is absolutely not mandatory to perform well. A shaded area in Figs.~\ref{fig:panel3} indicates the spreading of the results. There is none in plots \textbf{C} and \textbf{D}, meaning that all reservoirs perform exactly the same for a given $\sigma^\star$. Moreover, as $\sigma^\star$ is increased through the critical region, the dominant attractors change (see zooms in plot \textbf{C} and \textbf{D}), but surprisingly, there is no discontinuity in the performance. Indeed, even though the four attractor categories are present inside the gray area, their respective performance all align. This strongly suggests that the dynamics of the reservoir, as measured in the absence of input, is irrelevant for the performance. Only the value of $\sigma^\star$ matters. In both tasks, the average performance decreases monotonically with increasing difficulty via $\tau$ and $\delta$. In the memory task, we register a dip in performance close to the critical point. This goes against the common assumption that the edge of chaos is optimal for memory \cite{Natschlager2005}. In the prediction task, the peak of performance roughly coincides with the critical region, except for the greater difficulty, where the peak is slightly on the left. 

The picture is very different for $\sigma^\star > 0$ (Figs.~\ref{fig:panel3}\textbf{E} and \textbf{F}). First, the region in which some level of performance is observed is comparable to the critical region observed in free-running reservoirs. Second, there is substantial variability in performance across different reservoirs, as indicated by the large shaded areas. Despite this variability, there is an overall dependence of performance on $\sigma^\star$. The average performance of distinct dominant attractor categories is much noisier. However, despite being more noisy, the average performance of the distinct attractor categories aligns again, so there is still no evidence that the attractor category has any significant impact on the reservoir's performance. We observe that performance decreases as the difficulty of the memorization task increases, but interestingly, this trend appears to be inverted in the prediction task.

Once again, the two signs of  $\sigma^\star$ give rise to different behaviour. In particular,  networks with $\sigma^\star > 0$ memorize better and are less reliable than those with $\sigma^\star < 0$ but have poorer prediction capability. Yet, in all cases, attractors do not seem to be correlated to performance, as the top performance can be found in any of the four attractor categories.

\subsubsection{Discussion:}
Our results somewhat challenge the common assumption that the edge of chaos is optimal for performance and suggest that this is true for reservoirs with a majority of excitation but not necessarily with a majority of inhibition. Reservoirs with negative $\sigma^\star$ exhibit very reliable performances with very low reservoir-to-reservoir variability over a range in  $\sigma^\star$ much broader than the critical region. Since reservoirs behave the same, in practice, it is sufficient to generate one, with $\sigma^\star$ at the left of the critical region. However, if the goal is optimal memorization, it is wiser to choose $\sigma\sim0.4$ in the critical region and try different reservoirs until finding a good one, which requires training and testing.

\subsection{Cross-task performance} \label{p3:cross}
Beyond studying the performance in memorization and prediction separately, as often done (\cite{Bertschinger2004c}, \cite{Busing2010}), here we aim at answering the following question: are the reservoirs intrinsically good or bad, or does it depend on the task? In other words, are there general-purpose reservoirs and specialized ones? 

\subsubsection{Analysis:}
We analyze the absolute value of the performance of all reservoirs independently of the control parameter. For each reservoir, we study its performance in the memory task as a function of its performance in the prediction tasks (see section \ref{p3:performance} \textit{methodology}). For this, we fixed values of $\delta$ (memory) and $\tau$ (prediction), and we chose three levels of difficulty: simple ($\tau=5$, $\delta=-2$), average ($\tau=20$, $\delta=-6$) and difficult ($\tau=28$, $\delta=-10$).

\begin{figure*}
\centering
    \includegraphics[scale=0.6]{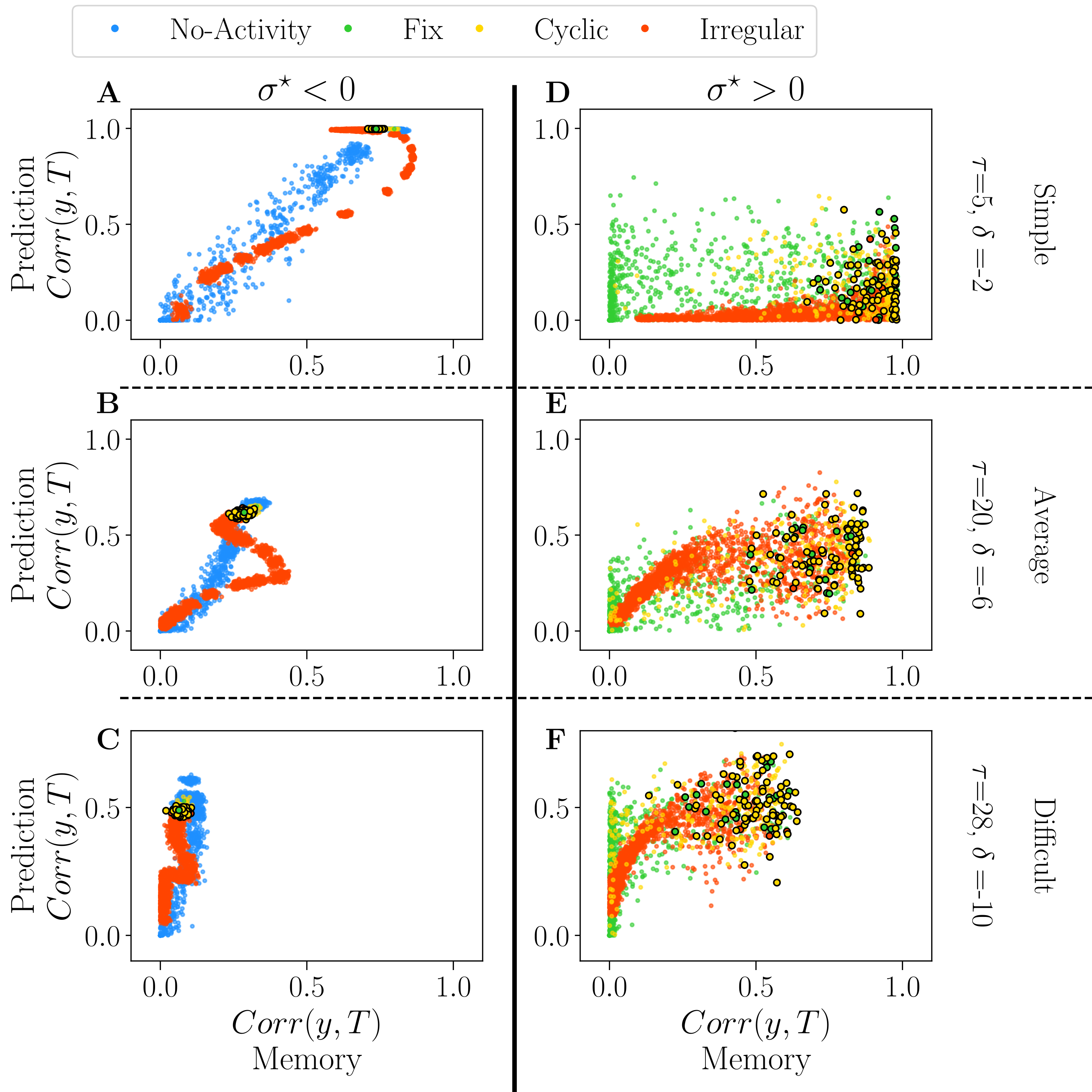}
    \caption{The performance of all reservoirs in the prediction tasks (Mackey-Glass) as a function of their performance in the memory tasks (white-noise), for $\sigma^\star<0$ (\textbf{A}, \textbf{B}, \textbf{C}) and $\sigma^\star>0$ (\textbf{D}, \textbf{E}, \textbf{F}). Reservoirs are again classified according to their dominant attractor (see supplementary material S\ref{suppm:stat_dominant}). Each dot represents an individual reservoir performance averaged over $5$ initial conditions. In all plots, dots with black edges display reservoirs taken at the critical regimes. We chose three different pairs of values for the parameters of the tasks, $\tau$ and $\delta$, each representing a different difficulty level: 1. Simple (\textbf{A} and \textbf{D}) the lowest difficulty in both tasks, $\tau=5$ and  $\delta=-2$; 2. Average (\textbf{B} and \textbf{E}) average difficulty for intermediary values of $\tau=20$ and $\delta=-6$; 3. Difficult (\textbf{C} and \textbf{F}) difficult task for higher values of $\tau=28$ and $\delta=-10$.}
\label{fig:panel4}
\end{figure*}

\subsubsection{Results:}
Fig.~\ref{fig:panel4} displays the performance of reservoirs as coloured dots. As in the previous section, each colour corresponds to the dominant attractor of the reservoir. The vertical axis represents the performance in the prediction task (Mackey-Glass), and the horizontal axis that of the memory task (white noise). The columns correspond to the sign of $\sigma^\star$, negative on the left, positive on the right. The rows correspond to the degree of difficulty, from simple (top) to difficult (bottom). 

For $\sigma^{\star}<0$ (left column), there is an apparent relationship in performance between the two tasks. The degree of difficulty roughly acts as a scaling factor on the curves, but the correlation is always clear. The reservoirs which are good at predicting do memorize well as well. There is, however, a reentrant region of irregular reservoirs (red) which are the best at memorizing but are not optimal for predicting, in particular for the intermediate degree of difficulty (see the red loop on Fig.~\ref{fig:panel4}\textbf{B}). This point corresponds to the maximum on the right of the dip observed in Fig~\ref{fig:panel3}\textbf{C}. Interestingly, in all difficulties, reservoirs at the critical point (encircled dots) create a narrow area with a good overall performance. This picture refines our previous analysis of the impact of attractors on performance, as it seems that extinguished reservoirs can perform slightly better at memory than the others, and the same for the irregular reservoirs at prediction. Nonetheless, the difference between attractors is rather small, and it could be argued that it is insignificant.

The picture is very different for $\sigma^\star>0$ (right column). Performances are distributed as clouds of points. In Fig.~\ref{fig:panel3}, we observe a large reservoir variability in both tasks. Some reservoirs are suitable for one task and bad for the other one. For intermediate and difficult tasks, some reservoirs outperform the best ones with $\sigma^\star<0$ in both memorization and prediction. Reservoirs at the critical point are found on the right of the plots, i.e. they promise good memorization but are nonetheless widely spread, especially in prediction (vertical axis). Overall, the distinct attractors occupy the space in overlapping and indistinguishable clouds; this confirms the previous analysis that attractors do not play a role in performance.

\subsubsection{Discussion:}
Correlations in performances are very different for both signs of $\sigma^\star$. Choosing a reservoir with $\sigma^\star\sim-0.66$ ensures a good, general-purpose reservoir but with suboptimal performance. In contrast, going into the positive side of $\sigma^\star$ may lead to the best reservoirs in a given task or even better general purpose reservoirs, but this comes at a price: those gems cannot be found by the statistical analysis we have performed on their free running activity.

\section{Conclusion} \label{conclusion}

One of the main issues in the field of RC is the lack of principled methodology \cite{Rodan2011} for reservoir design. This article aimed to quantify the impact of the random weight generation process to better understand the relationship between connectivity, dynamics, and performance. We demonstrated that the only control parameter is the ratio $\sigma^\star=\sigma/\mu$ through a Gaussian weight distribution, which indirectly regulates the excitatory/inhibitory balance. We found two critical points and observed that reservoirs typically possess a dominant attractor, regardless of their initial states.

We investigated the relationship between the performance, the control parameter, and the preferred attractor in memory and prediction tasks. Our results reveal that $\sigma^\star$, hence the excitatory-inhibitory balance $b$, has a strong impact on performance in the two considered tasks while the attractor dynamics have none. We showed how to select a control parameter region that ensures good performance, thus providing a very efficient way to obtain high-performance reservoirs. This region corresponds to high attractor diversity. For $\sigma^\star<0$, the critical region is narrower and does not necessarily coincide with the top of performance, while for $\sigma^\star>0$, the critical region corresponds to the performing region.

For the tasks, we showed that negative $\sigma^\star$ values produced superior results in prediction, with reliable performance and low reservoir-to-reservoir variability. Therefore, it is sufficient to perform free-running and pick a single value of $\sigma^\star$, preferably close to the critical point $\sigma^\star_c$. In contrast, positive $\sigma^\star$ values were found to have higher performance in memory tasks but with greater volatility. Since a given $\sigma^\star$ value can lead to diverse performance outcomes, generating random reservoirs and testing them during training to select the best performers is still necessary. Given enough trials, however, our findings suggest that $\sigma^\star>0$ can generate the bests general-purpose reservoirs.

\section{Future work} \label{future}

We tested the impact of dynamics on performance in two types of tasks: memory and prediction, for various time series. It would be interesting to test if and how the balance and attractor dynamic impact other types of tasks and inputs, notably classification, as it is also a standard task in machine learning. Moreover, extending the cross-task performance analysis to classification could potentially reveal interesting insights about its performance. 

Still, one surprising result is the limited impact of the intrinsic attractor dynamics on performance. One could test the robustness of this result by refining the attractor category, and future work may reveal greater performance sensitivity to attractor dynamics. For example, the extinguished category included all reservoirs with activities dying before $1000$ time steps. Refining the analysis could involve correlating performance with the average time before free-running reservoir activity dies out. Similarly, cyclic reservoirs could be refined by analyzing their period \cite{Kinoshita2009}, while some irregular activities may be considered cyclic when run for more extended periods. Moreover, it is possible that combining other types of analysis, such as correlation in space and time \cite{Metzner2022}, avalanche distribution size \cite{Siddiqui2018}, basins of attraction \cite{DelGiudice1998, Chinarov2000a, Kinoshita2009}, the number of attractors \cite{Cabessa2018}, and study of the reservoir topology \cite{Kinoshita2009, Masulli2016}, could provide better categorization of dynamics, with ultimately better predictive power of performance.

Finally, it would be of particular interest to see if our finding regarding the impact of the excitatory/inhibitory balance and dominant attractors also applies to other models, such as the quantum Ising spins, also used in the context of RC, which exhibit analogous phase transitions \cite{Martinez-Pena2021a}, and improved memory and prediction of time series in its vicinity \cite{Kutvonen2020}.

\section*{Ackowledgement}

EC extends gratitude to Matin Azadmanesh, Lucas Herranz, and Ismael Balafrej for critical reading of the manuscript. The authors also express their appreciation to their colleagues at NECOTIS for their valuable feedback and insightful discussions throughout the research process. Finally, the authors thank the reviewers for their constructive comments that helped improve the quality of the manuscript.

\section*{Conflict of Interest Statement}
The authors declare that the research was conducted in the absence of any commercial or financial relationships that could be construed as a potential conflict of interest.

\section*{Author Contributions}

This article represents the collaborative effort of EC, JR, and BR. EC conducted the research under the supervision of JR and BR. EC created the model, collected and analyzed the data, and wrote the manuscript. JR provided invaluable guidance and perspective throughout the process, along with BR, who also contributed to the writing and revision of the manuscript. All authors read and approved the final version of the article.

\section*{Funding}

This work was supported by the CRSNG/NSERC, the Canada First Research Excellence Fund, and the QsciTech program.

\section*{Code and data availability}

The research was conducted using a custom implementation. The authors created a repository \href{https://doi.org/10.5281/zenodo.8121795}{DOI:10.5281/zenodo.8121795} containing the code and data used to perform the analysis.

\bibliographystyle{apalike}
\bibliography{scibib.bib}

\section*{Supplementary Material}
\beginsupplement

\subsection{The experiments} \label{suppm:experiment}

This article proposes two sets of experiments: in the first setting, one is interested in the free time evolution of the reservoir without input ($u_i=0$). The second set is devoted to the measurement of the capacity of the reservoir to execute two tasks.

\subsection{Free running network} \label{suppm:nolearning}

Both experiments in this section are based on the same simulation and do not require the input and readout layers: 
\begin{enumerate}
    \item At time $t=0$ the network state is randomly initialized with provided \textit{seed}, and a fixed $20\%$ of active nodes ($x_i=1$). 
    \item Then the network is let run for a duration $D=2000$ time steps.
\end{enumerate}
\noindent

For each value of $\sigma^{\star}$, we generate 100 reservoirs, i.e. reservoirs having all the same architecture but different weights. We consider two ways to perform statistics.

\textbf{Statistics over many reservoirs} (section \ref{p2:activity}).

Each \textbf{reservoir} is run once in a \textit{free run simulation}. We report the result of $90$ values of $\sigma^{\star}<0$, $110$ values of $\sigma^{\star}>0$, and $80$ for $\mu=0$, for a total of $28,000$ simulations.

\textbf{Statistics of reservoirs} (section \ref{p2:dominant}).

Each \textbf{reservoir} is run $100$ times with different initial conditions. We report the result of $80$ values of $\sigma^{\star}$ for each sign, for a total of $1,600,000$ simulations.

\subsection{The control parameter} \label{suppm:sigma_values}

\subsubsection{Link between $\sigma^\star$ and $\rho$} \label{suppm:rho}

\begin{figure*}
    \centering
    \includegraphics[scale=0.4]{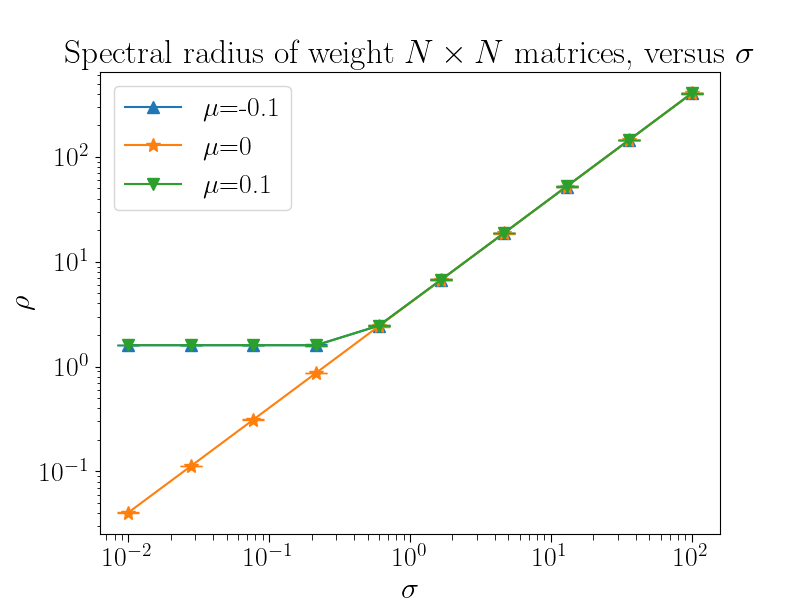}
    \caption{The spectral radius $\rho$, of Gaussian weight matrices, computed for different values of the mean $\mu(W)$, and as a function of the standard deviation $\sigma(W)$. Each dot represents the average computed on $10$ matrices, and errorbars represent two standard deviations.}
    \label{figs:1}
\end{figure*}

The spectral radius $\rho$ of a square matrix $W$ is the largest of its eigenvalues in absolute value. We will consider the case where the size $N$ of the network, here $N=10,000$ is large enough so that $\rho$ is self-averaging: it does not depend on the exact coefficients of $W$ but simply on the parameters $(\mu,\sigma)$ of their distribution. It is showcased in Fig.~\ref{figs:1} where the (inexistent) error bars represent the spread in the results for the calculation of $\rho$ for different reservoirs of identical $(\mu,\sigma)$. One can thus consider $\rho$ to be a function of $\mu$ and $\sigma$.

The property $\rho(\lambda W)=|\lambda|\rho(W)$ for any real $\lambda$ leads to $\rho(\lambda\mu,\lambda\sigma)=|\lambda|\rho(\mu,\sigma)$. 
In the case $\mu=0$, taking $\lambda=1/\sigma$ gives $\rho(0,\sigma)=\alpha|\sigma|$ with $\alpha=\rho(0,1)$ a parameter independent of the weight matrix. We thus find that while the dynamics of the network is independent of $\sigma$ for $\mu=0$ (see Fig.~\ref{fig:panel1}), $\rho$ can take any value. In this case, there is no link between $\rho$ and the dynamics of the network.

For $\mu\neq0$, taking $\lambda=1/\mu$ gives $\rho(\mu,\sigma)=|\mu|\beta(\sigma^\star)$ with $\beta(s)=\rho(1,s)$. Thus for a given value of $\sigma^\star$, $\rho$ can take any value. There is however a direct link between $\beta$ and $\sigma^\star$, shown in Fig.~\ref{figs:1}. For $\sigma=0$, the weight matrix is $W=\mu A$ with $A$ a matrix with $K$ ones and $N-K$ zeroes in each row, thus obeying  $\rho(A)=\beta(0)=K$, leading to $\rho(\mu, 0)=|\mu|K$, in agreement with our numerical simulation: $\rho$ saturates at low $\sigma$. For the limit of large $\sigma$ taking $\lambda=1/\sigma$ gives $\rho(\mu,\sigma)=\sigma\rho(1/\sigma^\star,1)\simeq\sigma\rho(0,1)=\alpha\sigma$, which is the result for $\mu=0$, as observed in Fig.~\ref{figs:1}.

While it may be tempting to replace $\rho$ by its renormalized value $\beta$, both quantities suffer from the same weakness: they do not depend on the sign of $\sigma^\star$, whereas the dynamics of the network strongly do.

\subsubsection{Control parameter values} \label{suppm:sigma_values}
Below are the parameter values for the experiment of free running (see methodology in \textbf{S}\ref{suppm:nolearning}):

\vspace{0.5cm}

\begin{minipage}[c]{0.4\textwidth}\leftskip=0.2\textwidth
    \begin{tabular}{||c |c |c||}
         Start & End & Step \\
         \cline{1-3}
         0.01 & 0.06 & 0.001 \\
         0.06 & 0.1 & 0.0007 \\
         0.07 & 0.1 & 0.001 \\
         0.1 & 0.2 & 0.01 \\
         0.2 & 1 & 0.1 \\
         1 & 11 & 1  
         \label{tab:mu_neg}
    \end{tabular}
    \captionof{table}{Values of $\sigma(W)$ for the first experiment, $\mu=-0.1$.}
\end{minipage}\hspace{0.2\textwidth} 
\begin{minipage}[c]{0.3\textwidth}
    \begin{tabular}{||c |c |c||}
         Start & End & Step \\
         \cline{1-3}
         0.01 & 0.07 & 0.001 \\
         0.07 & 0.2 & 0.01 \\
         0.2 & 0.5 & 0.02 \\
         0.5 & 1 & 0.1 \\
         1 & 11 & 1 
         \label{tab:mu_pos}
    \end{tabular}
    \captionof{table}{Values of $\sigma(W)$ for the first experiment, $\mu=0.1$.}
\end{minipage}
\vspace{0.2cm}

\subsubsection{Performance of tasks} \label{suppm:performance}

All experiments in this section test the performance in a task, where learning is required for the readout layer. For all tasks, the following framework is applied. Assuming that $F$ is the reservoir, the relationship between the input time series $u(t)$ and the output layer $y$ is $y(t)=F(u(t))$. The goal here is to learn the target $T(t)$, such that $y(t) = T(t)$ and the target is set to:
\begin{equation}
T(t) = u(t+\delta)    
\end{equation}

\noindent Here, $\delta$ is an integer which represents a number of time steps. This parameter serves the general purpose of setting the type of task:
\begin{itemize}
   \item \textbf{Memory task} for $\delta <0$, the reservoir output must reproduce the input received in the past.
   \item \textbf{Prediction task} for $\delta >0$, the network output must produce an input not yet seen by the reservoir.
\end{itemize}
The (integer) parameter $\delta$ also sets the difficulty of the task: the higher in absolute value, the more demanding it is.

\textbf{White noise memory} (section \ref{p3:main}). $u(t)$ is a zero mean, unit variance white i.i.d noise. Successive inputs are uncorrelated, so prediction is not involved, only memory. We report results for $\delta\in \{-14, -10, -6, -2\}$ (see figure \ref{fig:panel2}.\textbf{A} for an illustration). 

\textbf{Prediction of Mackey-Glass series} (section \ref{p3:main}).
$u(t)$ is the Mackey-Glass time series \cite{Hajnal2006}, which is common to benchmark reservoir computational capabilities. It is given by the following dynamical equation: 

\begin{equation}
    x_{t+1} = ax_t+\frac{bx_{t-\tau}}{c+x^d_{t-\tau}}
\end{equation}
\noindent We choose $a=0.9$, $b=0.2$, $c=0.9$, $d=10$ and $x_0=0.1$. The dynamic of this equation can be controlled by varying $\tau$: as $\tau$  increases, the time series evolves from periodic ($\tau=5$) to chaotic ($\tau=28$), with a continuous increase in complexity in between. Results are for a fixed $\delta=10$ and $\tau=\{5, 15, 20, 28\}$ (see figure \ref{fig:panel2}.\textbf{B} for an illustration). 

\subsubsection{Training} \label{suppm:training}
The two tasks follow the same protocol for training the readout:
\begin{itemize}
    \item The network receives the input $u$ for duration $D=2000$ time steps.
    \item The first $500$ time steps are discarded and considered transients. This value is empirically obtained, adapted from the signal at hand, and coupled to a convergence test for permanent regime detection.
    \item The training is performed on the following $1500$ time steps on the concatenated in-time reservoir outputs, using the optimization procedure described in \ref{met:learning}.
\end{itemize}
\noindent Each experiment consists in $40$ values of $\sigma^{\star}$, $100$ reservoirs per $\sigma^{\star}$ value, and each network is run $5$ times with different randomly tossed inputs (i.e., $40,000$ simulations). Each training is performed for $4000$ epochs (with a total of $640,000,000$ training epochs).

\subsubsection{Performance}

The metrics of performance are given by the \textit{Pearson correlation coefficient} between the output and the target (each of length 1500). A perfect match corresponds to a correlation of $1$ while $0$ means an output of the network is not better than random.

\vspace{0.5cm}

\subsection{Measure of networks dynamics} \label{suppm:measure}

To evaluate the reservoir dynamics, we used the most straightforward measure one can imagine: the \textit{activity}, which we defined as the sum of all spikes at each given time step, see supplementary material \ref{suppm:metrics}. We will take some time to make the reader appreciate the usefulness of such an approach. The sum of binary spikes is mathematically equivalent to how the magnetic field of spins in the ISING models \cite{Ising1925} is computed. In ISING, the electron spins are modelled as miniature magnets that can take binary values $[-1, 1])$, which flip with some probability depending on temperature and the coupling with other spins. The ISING model is one of the closest physical models to neural networks. In fact, it has been widely used to design neural networks, for example, in the famous Hopfield network \cite{Hopfield1982} or the Boltzmann machine \cite{Sherrington1975}. So one could argue that our activity signal is close to what Magnetoencephalography (MEG) is for brain activity \cite{Hamalainen1993}. 

Since this analysis does not depend on the micro-constituents, we make the case that the methodology performed in this article is easily transferable to other neural models and to other fields, such as neurosciences, physics, and the study of complex dynamical systems.

\noindent\textbf{Activity}: we define $A(t)$, the averaged activity of the network at time $t$, and for simplicity we will refer to it as \textit{activity}. It is the normalized sum of all neural states $x_i(t)$:

$$A(t)= \frac{1}{N}\sum_{i=1}^N x_i(t)$$
\noindent\textbf{Steady state of reservoir}: In the main text (part \ref{p2:activity}), we defined the steady state as $A$ for ease of notation, here for the sake of clarity, we define the steady activity $A_s$ of free-running reservoirs during $D=2000$ time steps, as $A_s=A(t \geq D/2)$, the activity of the last $D-D/2=1000$ time steps of free running simulations. 

\subsection{Metrics} \label{suppm:metrics}

\subsubsection{Permanent regime statistics:}
here are two statistical analyses we perform on the measure of the steady state activity $A_s$ of a given reservoir, where $\overline{(.)}$ represents the time-average over a quantity, and $\delta A(t)=A(t) - \bar{A}$: 

\begin{itemize}
    \item Permanent regime time-average:  $$\bar A_s =\frac{2}{D}\sum_{t=D/2}^{D} A(t)$$
    \item Permanent regime time-variance: $$\overline{\delta {A_s}^2}=\frac{2}{D}\sum_{t=D/2}^D(A(t)-\bar A_s)^2$$
\end{itemize}

\noindent\textbf{Average over reservoirs of permanent regime statistics}: in the \textit{first experiment} (see results in section \ref{p2:activity}), for each value $\sigma^{\star}$, we compute the average (over reservoirs generated with $100$ different seeds) of the permanent regime \textit{average} and \textit{variance} (overtime  time step $t$). Here $A^r_s$ denotes the stationary activity of the reservoir of index $r$ generated with $seed=r \times 100$, for a total number of reservoir $R=100$. We compute the average over reservoirs of the permanent regime statistics, where the back $\langle . \rangle$ denotes the average over reservoirs:
\begin{itemize}
    \item Average over reservoirs of the permanent regime average: $$\langle \bar A_s \rangle = \frac{1}{R}\sum_{r=1}^{R} \overline{A_s^r} $$
    \item Average over reservoirs of the permanent regime variance: $$\langle  \overline{\delta{A_s}^2} \rangle = \frac{1}{R}\sum_{r=1}^{R} \overline{(\delta A_s^r)^2}$$
\end{itemize}

\noindent\textbf{Statistics of the BiEntropy:} the BiEntropy solves one important limitation of the Shannon entropy, regarding the evaluation of binary words: as an example, let us take these two binary words "01010101" and "10110010". Although one is fully periodic and the other is somewhat random, they both have the same probability of occurrence of 1's and 0's; hence they have the same Shannon entropy. The Shannon entropy is thus insufficient to inform on the regularity versus the disorder of binary words. In contradiction, the binary entropy has been designed to discriminate patterns that the Shanon entropy could not \cite{Blackledge2020}.
We argue that in our context this metric is perfectly fitted to evaluate the binary words of spike time patterns. Since phase transition in such systems are known to be between an ordered and disordered phase, while the critical regime, also known as the edge of chaos, is supposedly a mixture of both. One property of particular interest to us regarding that metric, is that it is bounded between between 0 (for perfectly ordered words), and 1 (totally disordered).\astfootnote{For example, the word ”01010101” has $H_b=0.0078$, and the word ”10110010” has $H_b=0.7596$. A convenient way to discriminate both words.}

\begin{itemize}
    \item BiEntropy of permanent regime: the BiEntropy $H_b$ is computed on binarized steady activity $\delta A_s$. First, we subtract 
    the mean $A_s-\bar A_s$, and then all positive values are clamped to one and negative values to zero, resulting in a binary sequence. The $H_b$ is then computed on this sequence after converting it to a string.
    \item Average over reservoirs of the permanent regime BiEntropy: 
    $$\langle H_b \rangle = \frac{1}{R}\sum_{r=1}^{R} H_b^r$$
    \item Variance over reservoirs of the permanent regime BiEntropy: 
    $$\langle  \delta{H_b}^2 \rangle = \frac{1}{R}\sum_{r=1}^{R} (H_b^r-\langle H_b \rangle)^2$$
\end{itemize}

\noindent NB: given that neuron states are binary, one could wonder why employing a binarized version of a continuous variable like $A$. The reason is twofold: 1. the number of neurons in the network is $N=10000$. Therefore, computing $H_b$ on all neurons would be extremely costly. As such, reducing the number of neurons would be crucial, though it would necessitate a criterion for selection. This poses the risk of missing important information or introducing biases.  2. This would not be applicable in many real-world applications where access to the micro-constituents of the reservoir is difficult or even impossible. As such, we ensure our methodology is easily transferable to other areas comprising non-invasive studies of the brain.  


\section{Classification of attractors} \label{suppm:classification}

In this section, we provide a more grounded explanation of the choice of attractor categories. In the main text (part \ref{p2:dominant}), we developed a scheme to categorize activities by their respective attractor. Fig.~\ref{fig:panel2} showed the histograms of attractors as a function of $\sigma^\star$, and in this section, we provide a refined version of these statistics by adding two more categories of attractor:
\begin{itemize}
    \item \textit{Saturated attractors}: The steady activity is saturated when all neurons are active at all times.
    \item \textit{Non-trivial}: Any signal whose category of dynamics changes over time and that does not fit in previously mentioned types is considered non-trivial. This comprises cases where during the time window considered, activity suddenly changes from one type of dynamic to another. 
\end{itemize}

\begin{figure*}
    \centering
    \includegraphics[scale=0.5]{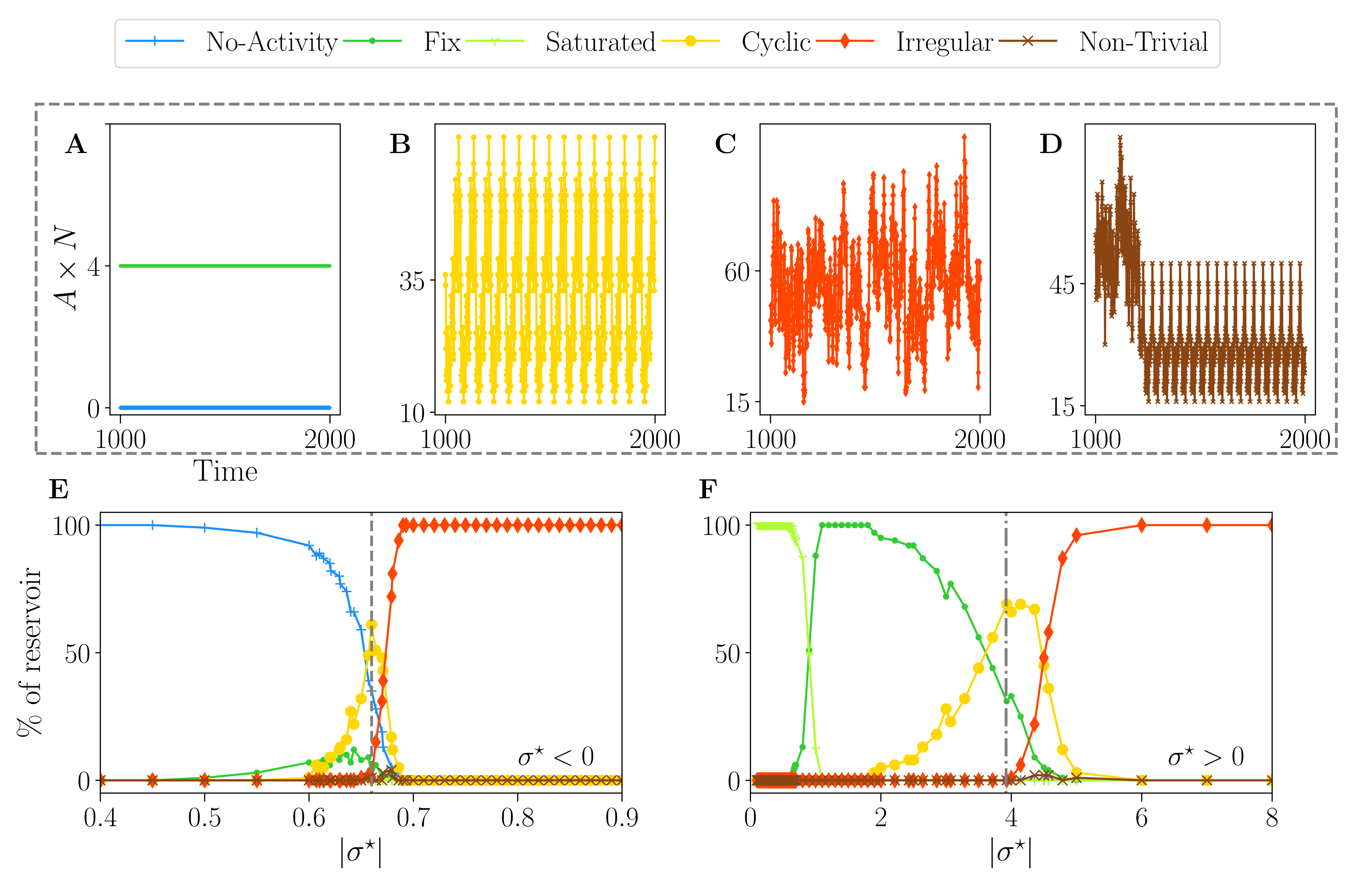}
    \caption{Examples of activities, classified into their respective category of attractor: (\textbf{A}) no-activity (\textcolor{dead}{$+$}), fix (\textcolor{fix}{$\sbullet$}); (\textbf{B}) cyclic (\textcolor{cyclic}{$\bullet$}), (\textbf{C}) irregular (\textcolor{chaos}{$\blacklozenge$}), and (\textbf{D}) non-trivial (\textcolor{ns}{$\times$}). The percentage of trials (or activities) belonging to each category of attractors is plotted against the control parameter $\sigma^{\star}$. (\textbf{A}) the phase transition for $\sigma^{\star}<0$, and (\textbf{B}) the phase transition obtained with $\sigma^{\star}>0$. In both cases, we zoom on the critical regions since the ordered, and irregular regimes have constant statistics. \textbf{E} and \textbf{F}: the statistics of attractors of free running simulations are computed for each value of $\sigma^{\star}$, for which $100$ reservoirs are run once (see methodology \ref{suppm:nolearning}). (\textbf{E}) $\sigma^\star<0$, and (\textbf{F}) $\sigma^\star>0$.}
\label{fig:panel2s0}
\end{figure*}

\noindent We show in Fig.~\ref{fig:panel2s0} examples of the various activities belonging to each category, No-Activity and \textit{Fix}  (\textbf{A}), \textit{Cyclic} (\textbf{B}), \textit{Irregular} (\textbf{C}), and lastly, \textit{Non-Trivial} (\textbf{D}). In the following paragraphs, we explain in more detail some specificities of the \textit{fixed} point attractor category and the reason why did not treat the \textit{Non-trivial} case separately from the irregular ones. 
First, we must mention that the \textit{fixed} point attractor category could, in theory, encompass the \textit{extinguished} and \textit{saturated} cases as well. This is because they all fit inside the definition of a time derivative of zero. The reason why we separated inactive reservoirs from the two others comes from percolation theory \cite{Coniglio1976}: formally, a reservoir has not percolated if the activity does not spread to infinity in space and/or time. While at the percolation threshold, activity will start to propagate indefinitely. We make the case that the percolation threshold, which does not coincide here with the critical points, constitutes another type of transition \cite{Cohen2010}, from inactive, to active, hence the distinction. This is visible in Fig.~\ref{fig:panel2s0}.\textbf{E}, as the fixed point attractors appear a bit before the \textit{cyclic} ones, around $\sigma^\star \sim 0.5$.

Second, in Fig.~\ref{fig:panel2s0}.\textbf{F}, we can see that the transition from \textit{Saturated} to \textit{Fix} attractor happens very early (in terms of $\sigma^\star$), compared to the rest of the attractors. So early, in fact, that it is not part of the phase transition. As you might recall, the analysis performed in part \ref{p2:activity} revealed that this region has a zero variance of both activity and BiEntropy. We conclude that the transition from \textit{Saturated} to \textit{Fix} is not related to a change of dynamics but only to a change in amplitude. As a result, we have chosen not to differentiate the two.

Thirdly, as one can clearly see in both Fig.~\ref{fig:panel2s0}.\textbf{E} and \textbf{F}, the \textit{Non-trivial} dynamics are very rare. As such, it is worth mentioning that in the analysis performed in \textbf{S}\ref{suppm:stat_dominant}, where we categorize reservoirs depending on their dominant attractors, not a single reservoir exhibits a dominance of \textit{Non-Trivial} dynamics. This is important because it means this category of attractors is irrelevant for finding correlations between dominant attractors and performance. 

\subsection{Diversity of reservoir attractor distributions} \label{suppm:entropy}

We compute the Shanon entropy $H_s$, with the goal of quantifying how varied are the attractor distributions of given reservoirs. Typically, if a reservoir activity always falls into one attractor, irrespective of the initial state, the Shanon entropy will give 0. On the other hand, the maximum entropy is obtained for a uniform distribution where each attractor is obtained $1/4$ of the time, and $H_s^{max}=-log(1/4)$. The Shanon Entropy is therefore normalized by its maximum value and averaged over the $100$ reservoir of each value of $\sigma^\star$. 

To quantify reservoir dynamics diversity, we plotted the average normalized entropy of reservoir attractor distributions, $\langle H_s/H_s^{max} \rangle$, against the control parameter for $\sigma^\star < 0$ (\textbf{A}) and $\sigma^\star > 0$ (\textbf{B}). The shaded area represents one standard deviation. Lower values of this quantity indicate a stronger dominance of one attractor. Outside of the critical region, initial conditions are irrelevant, and reservoirs always converge to a single attractor. The region where $\langle H_s\rangle$ increases correspond precisely to the region where the BiEntropy, shown in Fig.~\ref{fig:panel2s1}, is nonzero. In this region, different reservoirs may lead to different attractors. There is significantly more entropy, i.e. more competition between attractors, for $\sigma^\star < 0$. 

The higher variance in $H_s$ also implies increased reservoir-to-reservoir variability. This variability is more pronounced for $\sigma^\star < 0$, with the average entropy exceeding 0.25 near the critical point and the shaded area approaching 0.5. For $\sigma^\star > 0$, reservoirs generally exhibit stronger dominant attractors, as the standard deviation never surpasses 0.25. Despite these differences, the normalized entropy remains consistently below 0.5 for all values of $\sigma^\star$, irrespective of the sign. This value approximately corresponds to an attractor distribution where one attractor dominates 80\% of the initial conditions, indicating that most reservoirs possess a peaked distribution with a predominant attractor.

\subsection{Statistic of dominant attractors} \label{suppm:stat_dominant}

\begin{figure*}
    \centering
    \includegraphics[scale=0.5]{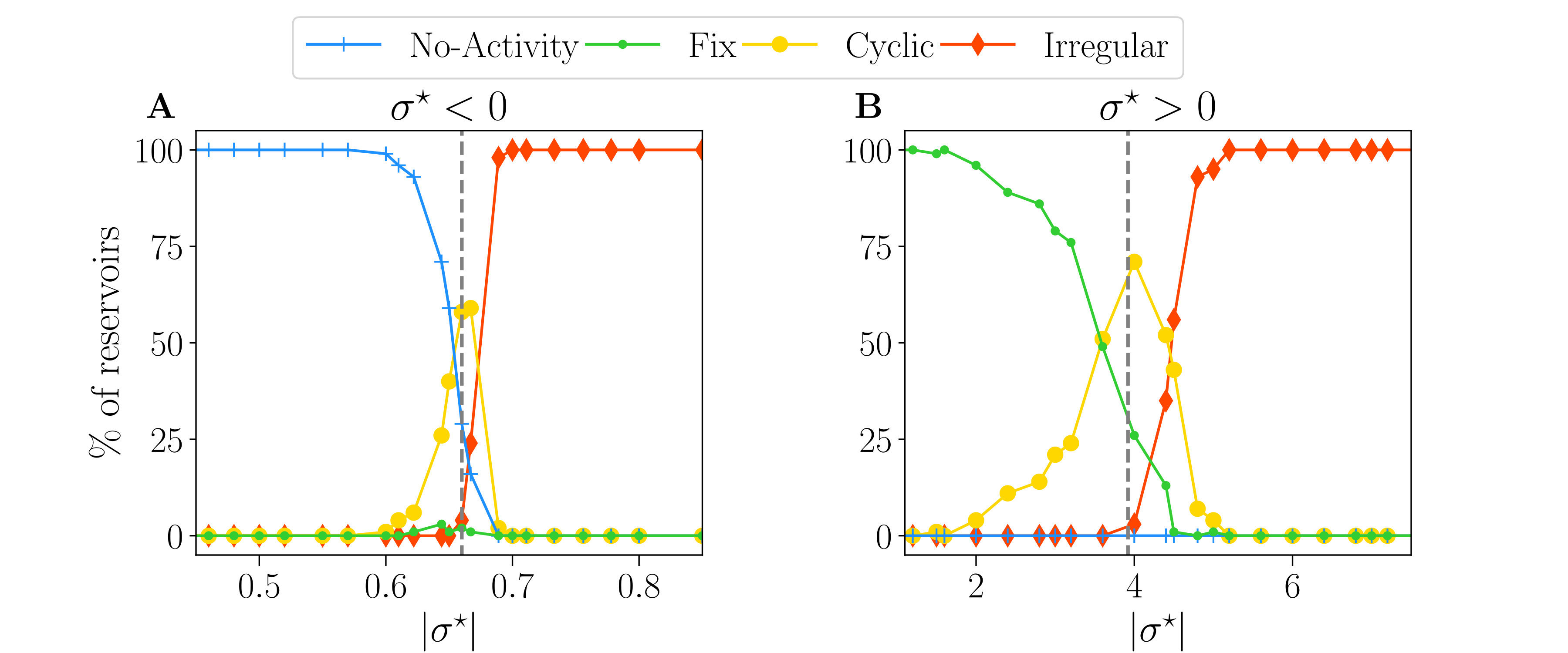}
    \caption{The statistics of reservoir dominant attractors. For each value of $\sigma^{\star}$, $100$ reservoirs are run $100$ times with different initial conditions (see \textbf{S}\ref{suppm:nolearning}). The resulting activities are classified into the category of attractors: no-activity (\textcolor{dead}{$+$}), fix (\textcolor{fix}{$\sbullet$}), cyclic (\textcolor{cyclic}{$\bullet$}), and irregular (\textcolor{chaos}{$\blacklozenge$}). The statistics of reservoirs with specified dominant attractor is plotted against the control parameter $\sigma^{\star}$. (\textbf{A}) $\sigma^{\star}<0$, (\textbf{B}) $\sigma^{\star}>0$.}
    \label{fig:panels3}
\end{figure*}

Fig.~\ref{fig:panels3} displays the statistics of the dominant attractors of each of the $100$ reservoirs generated by $\sigma^\star$ values. As one can note, the statistics are also unchanged from Fig.~\ref{fig:panel2}. Taken together, these results indicate strong statistical robustness, as averaging over reservoirs is almost equivalent to averaging reservoirs themselves. 

\end{document}